\documentclass[fleqn,usenatbib]{mnras}

\usepackage[flushleft]{threeparttable}
\usepackage{amsmath,lscape,epsfig}
\usepackage{amsfonts}
\usepackage{amsmath}
\usepackage{slashed}
\usepackage{graphicx}
\usepackage{epstopdf}
\usepackage{color}
\usepackage[usenames,dvipsnames]{xcolor}
\usepackage{ulem}
\usepackage{comment}

\usepackage[T1]{fontenc}

\DeclareRobustCommand{\VAN}[3]{#2}
\let\VANthebibliography\thebibliography
\def\thebibliography{\DeclareRobustCommand{\VAN}[3]{##3}\VANthebibliography}

\usepackage{graphicx}	
\usepackage{amsmath}	

\newcommand{\kms}{km\,s$^{-1}$}

\newcommand{\oxygenIII}{[O\,III]\,$\lambda\lambda$4959,5007}
\newcommand{\ironX}{[Fe\,X]\,$\lambda$6374}
\newcommand{\oxygenI}{[O\,I]\,$\lambda$6300}
\newcommand{\nitrogenII}{[N\,II]\,$\lambda\lambda$6548,6583}

\newcommand{\sulfurII}{[S\,II]\,$\lambda \lambda$6716,6731}

\newcommand{\oline}{[O\,III]$\lambda$5007}

\title[Gas emission of dwarf AGN candidates]{Spatially resolved spectroscopic observations of gas emission in dwarf galaxies hosting accreting black hole candidate}

\author[K. F. Heckler et al.]{
Kelly F. Heckler,$^{1}$\thanks{E-mail: kelly.heckler@acad.ufsm.b}
Rogemar A. Riffel,$^{2}$
Tiago V. Ricci$^{2,3}$
\\
$^{1}$Departamento de F\'isica, CCNE, Universidade Federal de Santa Maria, 97105-900 Santa Maria, RS, Brazil\\
$^{2}$Universidade Federal da Fronteira Sul, Campus Cerro Largo, 97900-000 RS, Brazil\\
}

\date{Accepted XXX. Received YYY; in original form ZZZ}

\pubyear{2015}

\begin{document}
\label{firstpage}
\pagerange{\pageref{firstpage}--\pageref{lastpage}}
\maketitle

\begin{abstract}
Recent studies on dwarf galaxies reveal that some of them harbor a massive black hole (BH), which is believed to have a similar mass of the Supermassive BH “seeds” at early times. The origin and growth of the primitive BHs are still open questions, since these BH seeds are hardly observed at high redshifts. Therefore, massive BH of dwarf galaxies can be the perfect candidates to untangle BH “seeds” properties and their influence on their host galaxy evolution, since massive BH may preserve their initial conditions due to its quiet merger and accretion histories.
We use optical integral field unit observations, obtained with the Gemini GMOS-IFU, to study the gas emission and kinematics in four dwarf galaxies, candidates to host massive BH, based on the analysis of their [Fe\,X] luminosities measured from SDSS spectra. The [Fe\,X] emission line is not detected in our GMOS in any of the galaxies, prompting speculation that its absence in our recent data may stem from a past tidal disruption event coinciding with the observation period of the SDSS data. 
All galaxies exhibit extended gas emissions, and the spatially resolved emission-line ratio diagnostic diagrams present values that suggest AGN photoionization from the [S\,II] -- BPT diagram. The gas velocity fields of all galaxies are indicative of disturbed rotation patterns, with no detection of gas outflows in any of the sources. 
Although the [S II] -- BPT diagrams indicate AGN photoionization, further confirmation through multi-wavelength observations is required to validate this scenario.

\end{abstract}




\section{Introduction}

There is no doubt about the presence of supermassive black holes (SMBHs), with masses in order of $M_{\rm BH} \sim 10^6 - 10^{9}$ M$_\odot$, at the centers of massive galaxies \citep{1995ARA&A..33..581Kormendy,1998AJ....115.2285Magorrian}. They are present in virtually all nearby galaxies with well-defined bulges and their growth seems to be linked to the host galaxy \citep{2013ARA&A..51..511Kormendy}. However, defining the complete range of phenomena associated with SMBHs still proves somewhat elusive. One of the central challenges lies in comprehending how SMBHs form and evolve, as well as in understanding their influence on the early stages of host galaxy evolution. This challenge is intricately tied to the obscured history of SMBH formation and growth in massive galaxies, shaped by a complex interplay of numerous mergers and accretion events \citep{2010MNRAS.408.1139VanWassenhove}.

Even less known and explored are the massive black holes (MBH), with masses of order of M$_\bullet \sim 10^4 - 10^6$ M$_\odot$ \citep{2020ARA&A..58..257Greene} which reside at the centers of low-mass galaxies (M$_\star \leq 10^{10}$ M$_\odot$) without bulges, such as dwarf galaxies in the Local Universe. 
These low-mass galaxies have a quiet merger history \citep{2011ApJ...742...13Bellovary}, hosting MBH that may still preserve many of their initial features (reminding BH "seeds"), making them perfect laboratories to investigate MBHs and early stages of galaxy evolution.

The evolution and growth of galaxies are mainly regulated by feedback mechanisms arising from Active Galactic Nuclei (AGN) and/or supernova (SN) explosions \citep{2012ARA&A..50..455Fabian, 2017NatAs...1E.165Harrison}.
Feedback processes are capable to shape galaxy growth by preventing (or promoting) stars to form by heating (or cooling) the gas in and outside the host galaxy \citep{2008ARA&A..46..475Ho, 2009Natur.460..213Cattaneo, 2012ARA&A..50..455Fabian}.
Moreover, feedback is the ingredient that was missing in galaxy evolution models to explain the galaxy luminosity function, solving the problem that the models were predicting more massive galaxies than observed in the Universe \citep{2005Natur.433..604DiMatteo, 2006ApJS..166....1Hopkins, 2012RAA....12..917Silk}. 
Outflows triggered by nuclear activity are extensively studied in nearby galaxies, and it is observed that they interact with the multi-phase gas \citep{2017ApJ...834...30Fischer, Forster_2019,Lutz_20,Bianchin_22,U_2022,Juneau_22,Rogemar_2023_kin} within the interstellar medium (ISM) through mechanical and radiative mechanisms, such as winds, radiation pressure, and jets \citep{2012ARA&A..50..455Fabian}, across galaxies of all masses \citep{2017NatAs...1E.165Harrison,  2018MNRAS.473.5698Dashyan,Penny_18,2019ApJ...884...54ManzanoKing}.
Moreover, external factors, such as mergers, and secular processes can also affect the ISM by disturbing the gas and triggering the feeding of the central BH or promoting the formation of new stars \citep{2010MNRAS.408.1139VanWassenhove}.
In this sense, detailed analysis of the ISM in galaxies containing an AGN is essential to disentangle which mechanism is taking place on the host galaxy, how the galaxy growth is shaped by these processes, and at what scales it occurs. 

AGN feedback is also thought to be important in early times, where the first galaxies were formed. Models suggest that AGN radiative feedback may have suppressed the growth of black hole seeds \citep{2012ApJ...754...34Jeon}. Thus, a detailed study in nearby dwarf galaxies may provide us clues to understand how the BH growth was shaped by feeding and feedback processes. This type of study is usually focused on massive galaxies while the detection of BH activity in dwarf galaxies is still quite recent \citep[e.g.][]{Mezcua20,Mezcua24,Kimbrell23}. In fact, AGN-driven outflows have been observed in some dwarf galaxies \citep{2019ApJ...884...54ManzanoKing}, and a recent study showed that AGN winds may have triggered star formation in the nearby star-forming galaxy Henize 2-10, suggesting a positive AGN feedback scenario \citep{2022Natur.601..329Schutte}.

The limited number of spatially resolved studies focused on active dwarf galaxies highlight the challenges in detecting this particular type of emission. Some studies reveal that the traditional method used to identify nuclear activity in the optical spectrum (like BPT Diagrams \citep{bpt_1981}) proves ineffective for most of dwarf galaxies \citep{2006MNRAS.371.1559Groves, 2019ApJ...870L...2Cann, 2021ApJ...922..155Molina_sample}. 
This is due to the fact that emission line ratios in the majority of these systems are dominated by star formation, combined with the low metallicity gas, leads to lower line ratio values for [N\,II]/H$\alpha$ and [O\,III]/H$\beta$, in BPT diagram \citep{bpt_1981, 2004ApJ...613..898Tremonti, 2013ApJ...774..100Kewley}. 
However, there are other mechanisms to identify nuclear activity in these objects, and a recent study of dwarf galaxies from the Sloan Digital Sky Survey (SDSS) reveals a sub-sample of active dwarf galaxies through the detection of the [Fe{\, \sc x}]$\lambda$6374 coronal emission line \citep{2021ApJ...922..155Molina_sample}. This coronal line is a known tracer of AGN activity, particularly in cases where its luminosity exceeds what can be explained by stellar phenomena.

In order to investigate the AGN influence on low-mass regime, we selected a sample of four nearby dwarf galaxies (\textit{z} $\sim 0.05$) with the highest luminosity of the [Fe\,X]\,$\lambda$6374 coronal emission line ($10^{36} - 10^{39}$ erg s$^{-1}$)  in the sample of \citep{2021ApJ...922..155Molina_sample}. 
This sample of galaxies exhibit intense optical emission lines, but no broad-line region (BLR) associated to a broad line component was detected for any of the four galaxies \citep{2021ApJ...922..155Molina_sample}. Using optical integral field spectroscopy, we map the gas emission line flux distributions and kinematics and investigate the origin of the gas emission in these galaxies. 

The paper is organized as follows. In Section \ref{sec:data} we describe our galaxies sample selection criteria, our observations, the data reduction procedure and measurements. Our results are presented in the Section \ref{sec:results} and discussed in Section \ref{sec:discussions}. Conclusions are presented in Section \ref{sec:conclusions}.

\section{Data} \label{sec:data}

A study developed by \citet{2021ApJ...922..155Molina_sample} with dwarf galaxies observed with the \textit{Sloan Digital Sky Server} (SDSS) survey, reveals the presence of active galactic nuclei by the detection of the coronal \ironX\, emission line. Although \ironX\, can also be produced by supernova explosions, the high luminosities  observed for this emission line ($L > 10^{36}$ erg s$^{-1}$)  can only be explained by the emission of gas photoionized by an AGN, as argued by these authors. We selected four AGN host candidates in dwarf galaxies, from the sample of  \citet{2021ApJ...922..155Molina_sample}, to perform spatially resolved observations of the gas emission structure and kinematics in their central regions. We selected nearby galaxies ($0.0296\leq z \leq 0.0544$) that follow the stellar mass constrain from \cite{2013ApJ...775..116Reines}, with stellar masses $M_\star \leq 3 \times 10^9$ M$_\odot$, that present the highest \ironX\ luminosities ($10^{38} - 10^{39}$ erg s$^{-1}$) in the sample of  \cite{2021ApJ...922..155Molina_sample}. Galaxy position on the sky and redshifts as well as the observation properties are presented on Table \ref{tab:observation}. Our sample is composed of the following galaxies: SDSS J030903+003846, SDSS J033549-003913, SDSS J033553-003946 and SDSS J225236-003317.

\subsection{Observations and data reduction}
\label{sec:datareduction}

The observations were performed with the Gemini North and South telescopes, depending on the position of the target on the sky. The Gemini North observations were conducted on June 10, 2023, under Program ID GN-2023A-Q-320, while the Gemini South observations were done on November 1, 2, and 14, 2022, as part of Program ID GS-2022B-Q-320.
We used the Integral Field Unit (IFU) of the Gemini Multi-Object Spectrograph (GMOS) \citep{AllingtonSmith_2002} in all observations with the one-slit mode, which results in data cubes covering a field-of-view (FOV) of $3.5 \times 5.0$ arcsec$^2$. The galaxies SDSS J030903+003846, SDSS J033549-003913 and SDSS J033553-003946 were observed using the B600 grating centered on the 5800 \AA, covering the spectral range of 4220--7410 \AA. SDSS J225236-003317 was observed using the B480 grating centered on the 5750 \AA, covering a spectral range of 4010--7820 \,\AA. These specific gratings were chosen to cover the strongest optical emission lines commonly observed in AGN hosts and star forming galaxies: H$\alpha$, H$\beta$, \oxygenIII, \nitrogenII \, and \sulfurII.

\begin{table*}

\caption{\label{tab:observation} Dwarf galaxies sky position and data observation properties. }


\begin{tabular}{cccccccc}
        Name & $\alpha$ & $\delta$ & $z$ & Program ID &  Grating & Seeing (arcsec) & \textit{d} (Mpc) \\
        (1) & (2) & (3) & (4) & (5) & (6) & (7) & (8)\\
        \hline
        \hline
        SDSS J030903.87+003846.8 & 47.26618 & 0.646299 & 0.030086	 & GS-2022B-Q-320 & B600   & 0.93  &  128 \\
        
        SDSS J033549.23-003913.1 & 53.95515 & -0.65365 & 0.03067 & GS-2022B-Q-320 & B600   & 0.91  &  130 \\
        
        SDSS J033553.12-003946.6 & 53.97134 & -0.66295 & 0.03016 & GS-2022B-Q-320 & B600  & 1.10  & 128 \\
        
        SDSS J225236.35-003317.7 & 343.15161 & -0.554828 & 0.05438 & GN-2023A-Q-320 & B480   & $0.68 $  & 231 \\
        
        \hline
    \end{tabular}
    \\
      \scriptsize{\textbf{Note}. Column (1): Galaxies name. Column (2)-(3): Right ascension and declination in degrees of the SDSS spectroscopic fiber obtained at the SDSS Science Archive Server. Column (4): Redshift. Columns (5)-(6): Gemini Telescope Program-ID and grating used in each observation. Column (7): Observational seeing obtained from r band acquisition images (see Sec. \ref{sec:datareduction}). Column (8): Galaxy distance in Mpc using H$_0 = 70.5$ km s$^{-1}$ Mpc$^{-1}$. }
\end{table*}

It is important to mention here that the Gemini South telescope presented heating problems, leading to complete saturation of amplifier 5, along with the presence of bright narrow columns in amplifiers 3, 8, and 11\footnote{More information about this issue can be found in https://noirlab.edu/science/news/announcements/sci23037.}
The GMOS-IFU detector (Hamamatsu \citep{2016SPIE.9908E..2SGimeno}) is formed by three CCDs  with two small physical gaps of $\sim 40$ pixels between. Each CCD is formed by 4 amplifiers, resulting in a total of 12 amplifiers. 
Thus, our observations have been affected by this issue, resulting in no reliable data between the wavelength range from $5870$ to  $6190$ \AA\ for the galaxies observed with the Gemini South telescope.

The data reduction was performed using the {\sc gemini} package available on {\sc iraf}, encompassing various stages: bias and flat-field corrections, removal of cosmic ray artifacts with the LACOS algorithm \citep{van_Dokkum_2001}, and extraction of all observed spectra. Wavelength calibrations were performed using CuAr lamp spectra. For the flux calibrations, the standard star Hiltner 600 was observed as part of the GS-2022B-Q-32 programme and observations of the BD+28 4211 standard star were performed as part of the GN-2023A-Q-320 programme. All data cubes were created with a pixel size of 0.1 arcsec. After this conventional data reduction protocol, techniques detailed in \cite{menezes2019} were used to eliminate high- and low-frequency noise in both spatial and spectral dimensions from the data cubes. Additional information on the data reduction and processing procedures are presented in \cite{Ricci2014a} and \cite{menezes2019}, providing a more comprehensive understanding of the applied methods. The instrumental broadening, estimated by fitting emission lines present in the CuAr spectra, is $\sigma_{\rm inst}\approx$ 40\,km\,s$^{-1}$ for all galaxies.

\subsection{Measurements} \label{sec:measurements}

To map the flux distribution and kinematics of the ionized gas, we fit each emission line profile by using a single Gaussian function. To perform the fits, we use the {\sc ifscube} code \citep{daniel_ruschel_dutra_2020_3945237,Ruschel-dutra_2021}, a Python tool optimized for analysing IFU data. After a visual inspection, we note that the emission lines present a well-behaved profile, with no need for more than a single Gaussian profile to reproduce each line. 
We tied the kinematics of emission lines that are formed under the same environmental conditions to reduce the number of free parameters in the fit: the hydrogen lines (H$\alpha$, H$\beta$ and H$\gamma$, when present), the \oxygenIII, \nitrogenII, and \sulfurII\, doublets, by keeping fixed the velocity and velocity dispersion of each group of lines.
For \oxygenIII \, and \nitrogenII\, doublets, we also kept the flux ratio fixed to their theoretical values, of 2.98 for [O\,III]\,$\lambda$5007\,/\,[O\,III]\,$\lambda$4959 and 3.06 for [N\,II]\,$\lambda$6583\,/\, [N\,II]\,$\lambda$6548 \citep{Osterbrock2006}. 

The fit of the emission lines begins in the position corresponding to the peak of the H$\alpha$ emission line and proceeds in a spiral loop. As we do not detect significant stellar absorption features in the spectra of the observed galaxies, we do not fit the underlying stellar population contribution. However,  we fit a pseudo continuum with an one-degree polynomial to account for the underlying continuum emission.  

Despite these galaxies have been selected based on the \ironX\ luminosity from \citet{2021ApJ...922..155Molina_sample}, the visual analysis of the SDSS spectra clearly shows the detection of this emission line for SDSS J030903+003846 and SDSS J033549-003913. This line is marginally detected for SDSS J033553-003946 and is not present or very weak for SDSS J225236-003317.  In our GMOS spectra, we do not detect the \ironX\ in any galaxy. 
Thus, we will focus our analysis on the strong emission lines.

\section{Results}\label{sec:results}

\begin{figure*}
    \centering
    \includegraphics[width=\linewidth]{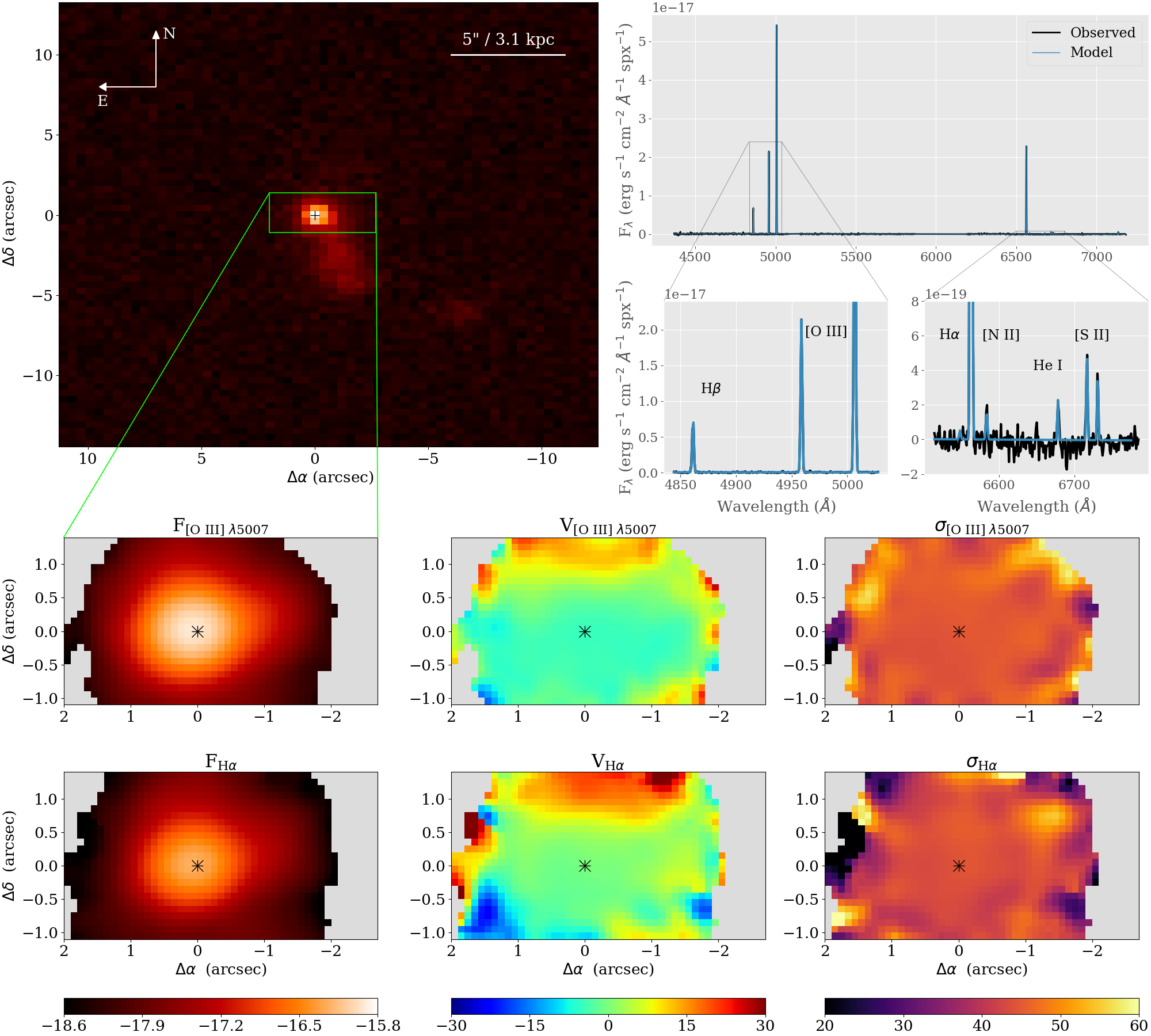}
    \caption{Results for SDSS J030903+003846. Left top panel: large scale SDSS g band  image of the galaxy obtained from the SDSS Science Archive Server (Data Release 12), with the GMOS IFU field of view represented by the green rectangle. Top-right panel: Spectrum of the spaxel corresponding to the location of the peak of the emission-line fluxes, which is co-spatial with the position of the continuum emission for this galaxy. The central and right panels in the second row show a zoom into the region of blue and red regions of the spectrum. The two-bottom rows show the flux distributions (left), velocity field (center) and velocity dispersion map (right) for the [O\,III]$\lambda$5007 (second bottom row) and H$\alpha$ (bottom row) emission lines. The color bars show the fluxes in logarithmic units of erg\,s$^{-1}$\,cm$^{-2}$\,\AA$^{-1}$ and the velocity and velocity dispersion values in km\,s$^{-1}$. The velocity field are shown after the subtraction of the systemic velocity of the galaxy. The $+$ sign mark the location of the nucleus, corresponding to the position of the peak of the continuum and the $\times$ sign marks the position of the H$\alpha$ emission peak. The gray regions correspond to masked locations where the emission lines are not detected with a signal-to-noise ratio of at least 3.}
    \label{fig:J030903_maps}
\end{figure*}

    \begin{figure*}
        \centering
        \includegraphics[width=\linewidth]{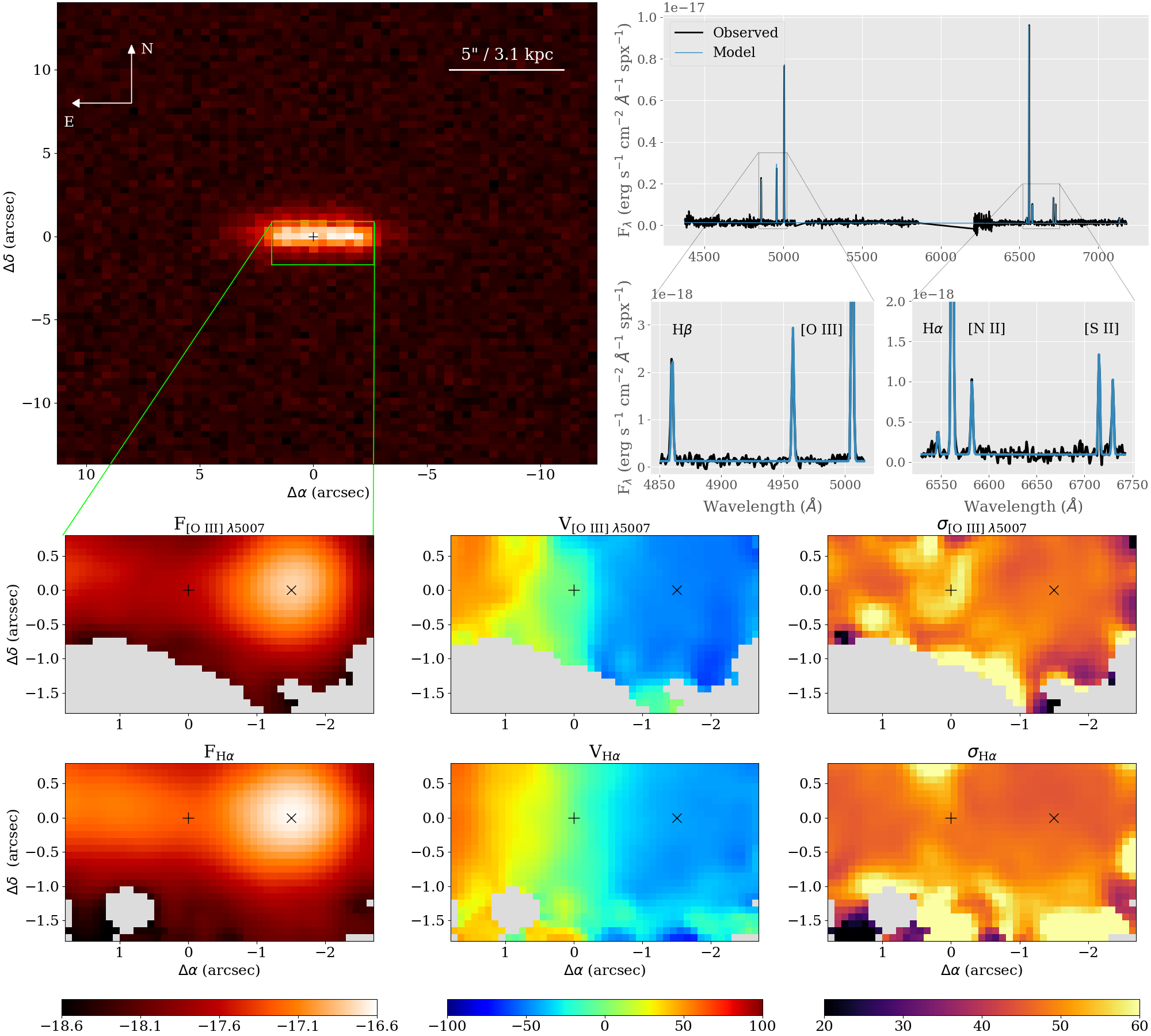}
        \caption{Same as Fig.~\ref{fig:J030903_maps}, but for SDSS J033549-003913. The black cross marks the H$\alpha$ peak emission and the black plus sign represents the galaxy nucleus.}
        \label{fig:J033549_maps}
    \end{figure*}

    \begin{figure*}
        \centering
        \includegraphics[width=\linewidth]{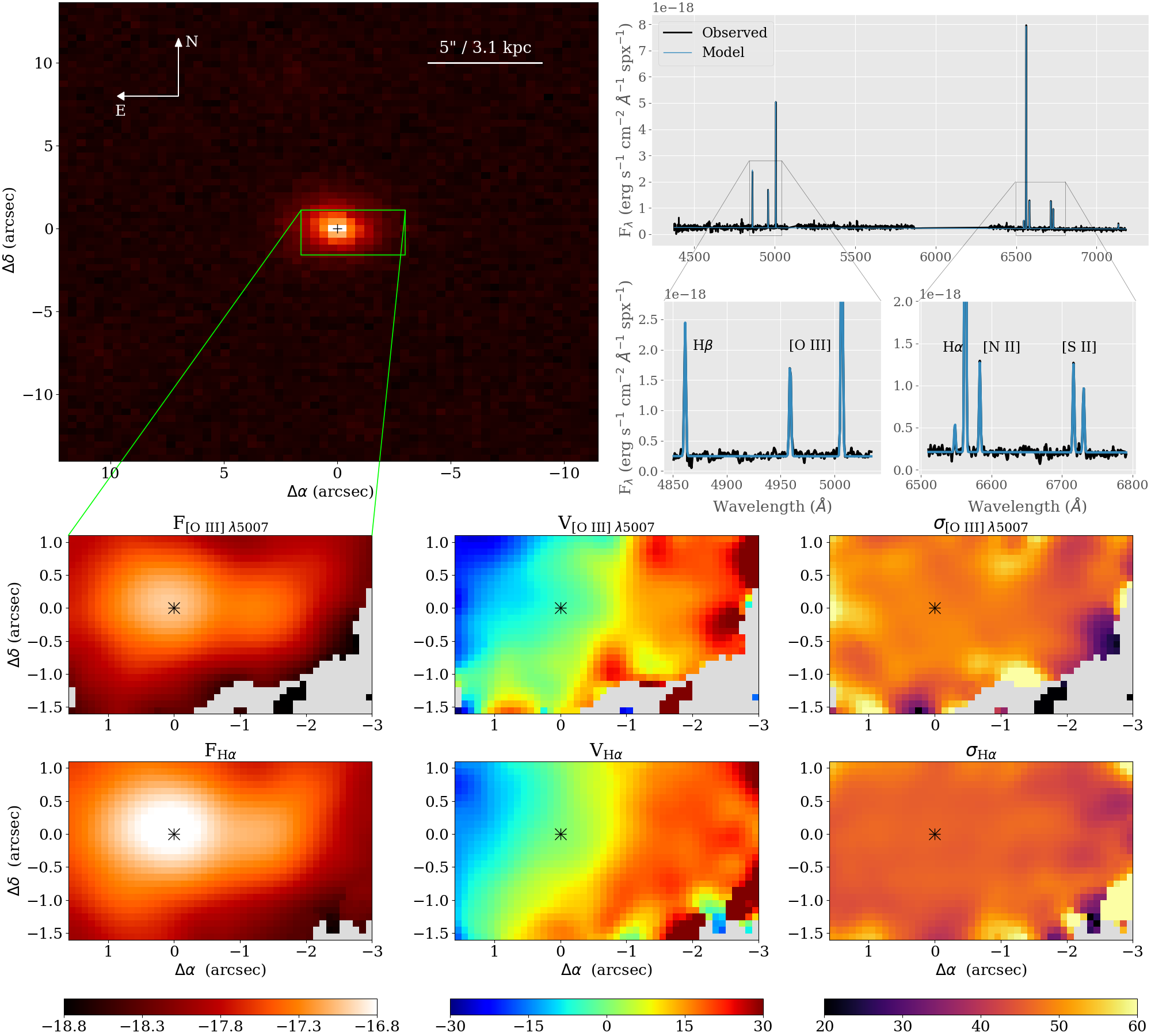}
        \caption{Same as Fig.~\ref{fig:J030903_maps}, but for SDSS J033553-003946.  The black cross marks the position of the nucleus (co-spatial with the line emission peak).}
        \label{fig:J033553_maps}
    \end{figure*}

    \begin{figure*}
        \centering
        \includegraphics[width=\linewidth]{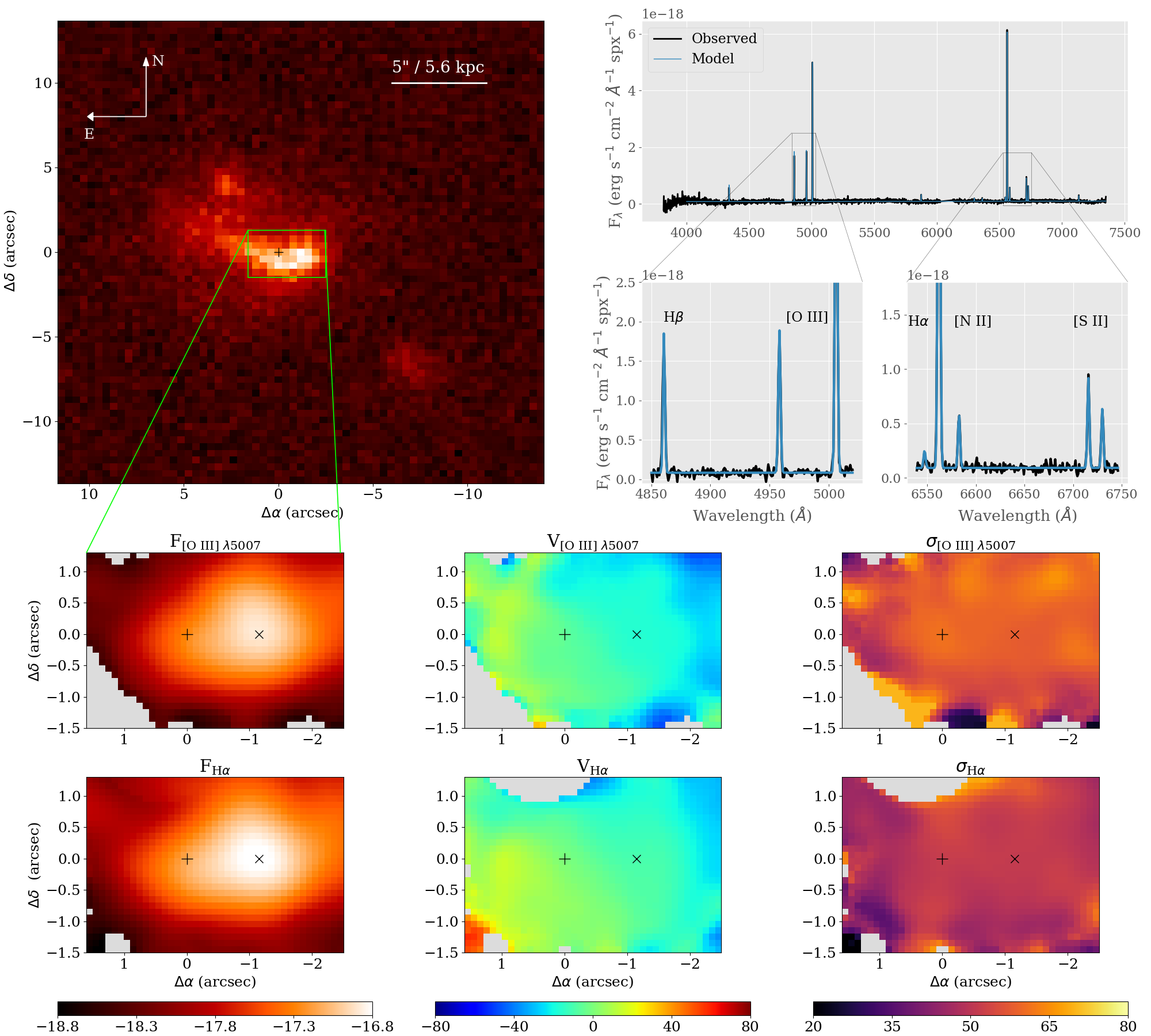}
        \caption{Same as Fig.~\ref{fig:J030903_maps}, but for SDSS J225236-003317. The black cross marks the H$\alpha$ peak emission and the black plus sign represents the galaxy nucleus.}
        \label{fig:J225236_maps}
    \end{figure*}

In Figures \ref{fig:J030903_maps}, \ref{fig:J033549_maps}, \ref{fig:J033553_maps} and \ref{fig:J225236_maps} we present the results for the four galaxies. For each galaxy, the SDSS large scale images in the g band (centered on 4686 \AA) is shown on the top left panel. These images were obtained from the SDSS Science Archive Server\footnote{Available on https://dr12.sdss.org/fields.} (DR12). The green rectangle represents the GMOS FoV. In the top right panel, we show the spectrum for the spaxel corresponding to the position of the peak of the H$\alpha$ emission (marked with the $\times$ sign in the maps) and a zoom in the blue ([O\,III]+H$\beta$) and red ([N\,II]+H$\alpha$ and [S\,II] lines) regions in the second row. The black line represents the observed spectrum and the blue line represents the modelled spectra. The two-dimensional maps for the [O\,III]$\lambda$5007 and H$\alpha$ flux distributions and kinematics are presented in the two bottom rows. The first column shows the flux distributions, the second column shows the radial velocity maps and the third column shows the velocity dispersion maps.  
The $+$ sign mark the location of the nucleus, corresponding to the position of the peak of the continuum and the $\times$ sign marks the position of the H$\alpha$ emission peak. 
The gray regions in these maps correspond to masked locations where the amplitude of the corresponding emission line is smaller than 3 times the standard deviation in the continuum in regions close to that line.  We show the maps for the other emission lines in Appendix~\ref{appendix}, for all galaxies. Below, we provide a detailed description of the results obtained for each object.

\subsection{ SDSS J030903+003846}

The peak of the continuum and emission line fluxes for this galaxy are cospatial. Its spectra show strong \oxygenIII\ and H recombination lines (Fig. \ref{fig:J030903_maps}). In addition, the \sulfurII\ and \sulfurII\ emission lines also present within the nuclear region ($<$ 1 arcsec). On the other hand, the low ionization emission lines of \nitrogenII\ and \oxygenI\ were not detected for this object. Results for the \oxygenIII\ and H$\alpha$ are shown in Fig.~\ref{fig:J030903_maps}, while results for the remaining emission lines are shown in Fig.~\ref{fig:1g_j030903}.

As observed in Fig.~\ref{fig:J030903_maps}, the \oline\ and H$\alpha$ flux distributions are extended over the whole GMOS field, with the highest flux levels revealing a slightly extended structure to the west of the nucleus. The gas velocity fields show a disturbed rotation pattern with a projected velocity amplitude of $\sim$30\,\kms\ and orientation of the lines of nodes approximately along the north-south direction, which is similar to the orientation of the extended emission structure seen in the large scale SDSS image (top-left panel), beyond the GMOS FoV. The velocity dispersion maps show values lower than 60\,\kms, with the highest ones observed in regions farther than 1 arcsec from the nucleus.

\subsection{SDSS J033549-003913}

For this galaxy, the peak of the emission-line flux distributions is observed $\approx$1.5 arcsec west of the nucleus. SDSS J033549-003913 presents extended emission of strong high ionization lines such as \oxygenIII\, and Hydrogen recombination lines (Figure \ref{fig:J033549_maps}). Low ionization emission lines were detect and they also present extended emission for the \nitrogenII\, and \sulfurII\, doublets. Emission of \sulfurII\, is restricted to a region with $\sim$0.8 arcsec centred at the position of the peak of the emission-line flux distributions. The maps for the \oxygenIII\ and H$\alpha$ are depicted in Fig.\ref{fig:J033549_maps}, whereas the results for the other emission lines are shown in Fig.\ref{fig:1g_j033549}.

As can be seen in the bottom right panel of Figure \ref{fig:J033549_maps}, the emission-line flux distributions are extended and more elongated along the east-west direction, following the large-scale shape of the galaxy (top right panel of the figure). The velocity field presents a clear rotation pattern with blueshifts observed to the west and redshifts to the east of the galaxy's nucleus and a projected velocity amplitude of $\sim$100\,km\,s$^{-1}$. The velocity dispersion maps show values $\sim$ 50 \kms\ at most locations.

\subsection{SDSS J033553-003946}

This galaxy presents extended high ionization emission lines such as \oxygenIII\, and H recombination lines and unresolved emission of \sulfurII. We also detect extended low ionization emission lines as \nitrogenII\, and \sulfurII\ (Figures~\ref{fig:J033553_maps} and \ref{fig:1g_j033553}). The flux distributions for all emission lines are extended along the east-west direction, except for the \sulfurII, which emission is unresolved and restricted to the nucleus. From the gas velocity fields, it can be observed that the nuclear region has velocity values near zero and a rotation pattern is revealed with positive velocity values to the south-west direction and
blueshift velocities at northeast. As for the other objects, this galaxy also presents small velocity dispersion values, with a mean value of $\sim$ 50 \kms.

\subsection{SDSS J225236-003317}

This galaxy presents extended emission over the whole GMOS FoV for most emission lines (Figs.~\ref{fig:J225236_maps} and \ref{fig:1g_j225236}). The gas emission is more extended along the east-west direction, following the morphology seen in the SDSS image, but with the peak of the line emission, displaced from the nucleus of the galaxy by $\sim$1.2 arcsec to the west. The gas velocity field show mostly blueshifts with the highest values seen in regions close to the borders of the GMOS FoV view to the west and to the north. Values nearly zero are seen in the inner $\sim$1 arcsec radius and some redshifts are seen to the east of the nucleus, suggesting a disturbed rotation pattern. Typical values of the velocity dispersion are $\sim$ 50 \kms, observed  across the entire FoV.

\section{Discussions} \label{sec:discussions}

\subsection{The non-detection of the \ironX\, emission line}

In this work, we present GMOS-IFU observations of four dwarf galaxies selected as AGN candidates from \cite{2021ApJ...922..155Molina_sample} sample. \citet{2021ApJ...922..155Molina_sample} selected dwarf galaxies (M$_\star \leq 3 \times 10^9$ M$_\odot$) with detectable emission of the \ironX\, coronal line from the SDSS. From that sample, we select galaxies with the highest \ironX\, luminosity ($L_{\mathrm{[Fe\,X]}} \geq 10^{38}$ erg s$^{-1}$), which was argued that only supernovae explosion was not enough to describe the observed luminosity, taking AGN as the possible ionization source.

\begin{figure*}
    \centering
    \includegraphics[width=\linewidth]{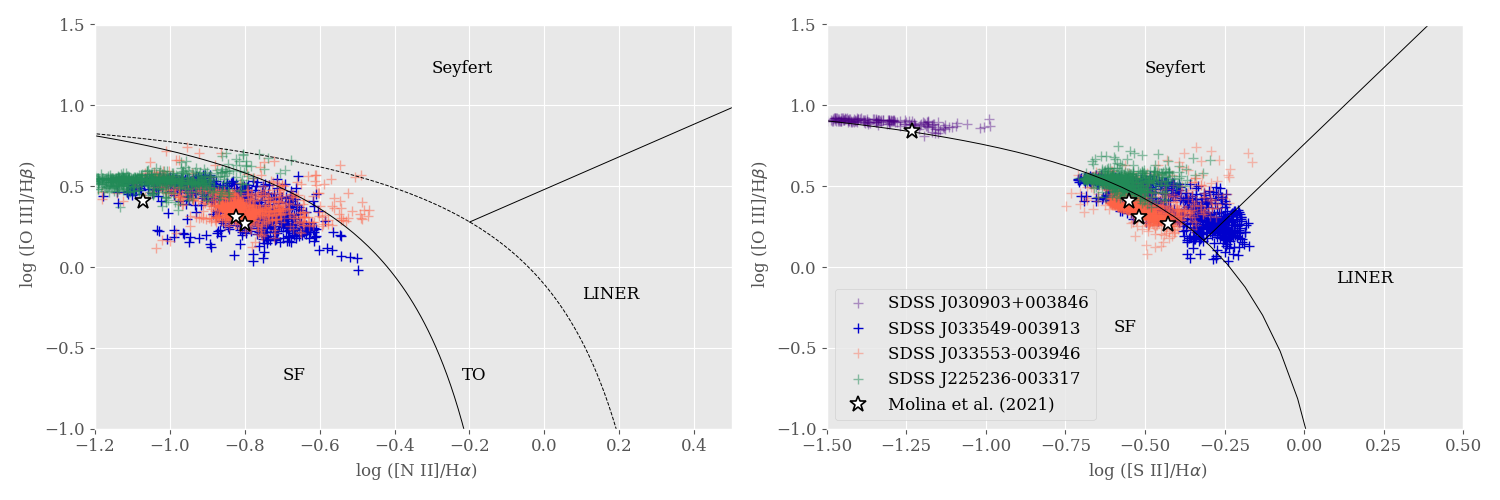}
    \caption[Optical BPT diagnostic diagrams \citep{bpt_1981} for our sample of galaxies.]{Optical BPT diagnostic diagrams \citep{bpt_1981} for our sample of galaxies. Left panel: [N\,II] -- BPT diagram. This diagram distinguishes between emissions arising from star formation (SF), AGN activity (Seyferts and LINERS - low-ionization nuclear emission-line regions), and transitional objects (TO), which may exhibit contributions from both star formation and AGN activity. 
    Right panel: [S\,II] -- BPT diagram. This diagram distinguishes between emissions due to star formation (SF) and AGN activity.
    Each color represents one galaxy from our sample: SDSS J030903+003846 is indicated by purple, SDSS J033549-003913 by blue,  SDSS J033553-003946 by orange, and SDSS J225236-003317 by green. The black stars represent the emission line ratio obtained from \citet{2021ApJ...922..155Molina_sample} for the same galaxies of our sample.}
    \label{fig:bpt}
\end{figure*}

The high ionization potential of \ironX\, $\sim 262.1$ eV \citep{kramida2022_NIST} requires a powerful event, such as supernovae explosions, BH activity or tidal disruption events (TDEs) \citep{2012ApJ...749..115Wang, 2023MNRAS.525.1568Short} to be produced. In that context, luminous \ironX\, emission has become a valuable tool for detecting low-mass BH activity in dwarf galaxies \citep{2021ApJ...922..155Molina_sample}.
Recent studies suggest that this coronal emission may be spatially resolved when associated with an AGN \citep{2021MNRAS.506.3831Fonseca-Faria, 2023ApJ...946L..38Reefe}, and this leads us to questioning the ionizing source of the \ironX\, emission, detected in SDSS and not detected in our work.

Initially, we first considered the possibility of a data-related issue for the non-detection of this emission line, since we see a pattern of non-detection in all galaxies from our sample. However, we disregarded that since we found that the J030903+003846 galaxy was also observed with MUSE in 2019, available at the ESO Database ('0104.D-0503(B)' ESO program identification), and by a detailed inspection of the galaxy spectra, we also do not detect any emission of \ironX\, on these data. 
Thus, there are some speculations that we can do about the non detection of \ironX. 
It is possible that the observed coronal emission on SDSS can be contaminated by Si II\,$\lambda$6371 leading to the misidentification \ironX\ emission line, as reported by \cite{2023RNAAS...7...99Herenz}. 

Other plausible explanation is a tidal disruption event. When a star passes within the tidal disruption radius \citep{1975Natur.254..295Hills} of a massive or supermassive black hole, the star is torn apart and forms a luminous flare that becomes part of the gas accretion disk that feeds the central black hole, and part of the energy is emitted in optical, UV and x-rays \citep{2021ARA&A..59...21Gezari}.  
So, TDEs also serve as probes to massive and supermassive black holes \citep{2020SSRv..216...32French, 2021ARA&A..59...21Gezari}, and there have been about 56 TDEs reported in the literature up to the year 2021 \citep{2021ARA&A..59...21Gezari}, including detections in dwarf galaxies \citep{2014ApJ...792L..29Maksym}. The TDE rate is linked to the size of the black hole \citep{1999MNRAS.309..447Magorrian, 2004ApJ...600..149Wang, 2016MNRAS.455..859Stone, 2018ApJ...852...72VanVelzen, 2021ARA&A..59...21Gezari}, as well as the light curve from the tidal flares \citep{2020SSRv..216...32French} with timescales that go from months to years \citep{2007A&A...462L..49Esquej, 2018NatAs...2..656Lin, 2021ARA&A..59...21Gezari}. 
These TDEs can ionize the gas in their surroundings, producing coronal lines such as [Fe X]\,$\lambda$6376 and [Fe XIV]\,$\lambda$5304, known as extreme coronal line emitters \citep{2012ApJ...749..115Wang}.

Not only TDEs, but SNe explosions are also capable of producing intense coronal lines, through their emission in soft X-ray, as mentioned before, and their luminosity can also luminosity can also disappear within several years. Therefore, SNe should also not be ruled out as a source of \ironX\, ionization. Considering that we selected galaxies with the highest \ironX\, luminosity from the sample of
\cite{2021ApJ...922..155Molina_sample}, with luminosities that were about 2 orders of magnitude higher than the luminosity of the brightest SNe, the origin of the [Fe X] emission in the SDSS spectra of the galaxies studied here may be less probable than the possibilities discussed above.

\subsection{The origin of the gas emission}

To identify the gas excitation mechanisms in these sources, we constructed the BPT diagnostic diagrams \citep{bpt_1981}. We present [N\,II] -- BPT and [S\,II] -- BPT based diagrams in the left and right panels of  Figure \ref{fig:bpt}, respectively. Each galaxy is represented by a distinct color: SDSS J030903+003846 is indicated by purple,  SDSS J033549-003913 by blue,  SDSS J033553-003946 by orange, and SDSS J225236-003317 by green.  SDSS J030903+003846 galaxy does not present  \nitrogenII\, emission and, thus, it is not included in the [N\,II] -- BPT  diagram.  The spatially resolved BPT-diagrams are shown in Figures~\ref{fig:bpt_j030903}--\ref{fig:bpt_j225236}. 

By analyzing the [N\,II] -- BPT diagram, most of the spaxels of the observed galaxies fall into the starforming region, with a few points falling into the transition region were both starformation and nuclear activity are able to ionize the observed gas. On the other hand, the [S\,II] -- BPT diagrams present values in the AGN region for all galaxies. Using the measurements from \citet{2021ApJ...922..155Molina_sample}, the four objects are classified as starforming galaxies, and only SDSS J030903+003846 would be classified as a low metallicity AGN in the [S\,II] -- BPT diagram, as can be seen in Fig. \ref{fig:bpt}. 
\citet{2021ApJ...922..155Molina_sample} also found that SDSS J033553-003946 presents near-infrared emission consistent with photoionization by AGN through the  near infrared (NIR) diagnostic diagrams. 
The SDSS line ratios shown in the figure are similar to the values that we obtained, however the integrated spectra has the limitation in distinguishing between star formation and nuclear activity since it is not possible to separate nuclear from the extra-nuclear emission within the aperture covered by the SDSS data.

Low stellar mass regime is dominated by low metallicity gas, that can affect emission line ratios behavior of narrow line regions (NLR). Nitrogen becomes more rare compared to hydrogen abundance as metallicity decreases, implying that the [N\,II]/H$\alpha$ emission line ratio values shift to the left in [N\,II] -- BPT diagram, falling predominantly into the starforming region even in a presence of a low-luminosity AGN \citep{2006MNRAS.371.1559Groves, 2006MNRAS.372..961Kewley, 2013ApJ...775..116Reines, 2019ApJ...870L...2Cann}.
Higher nebula temperatures and the average ionization state can increase [O\,III]/H$\beta$ line ratio indicating a strong ionization source, whose origin can be newborn stars or nuclear activity \citep{2006MNRAS.371.1559Groves}. While the [N\,II] -- BPT diagram may not be as accurate to identify the gas ionization mechanism in low metallicity regimes, we can use the [S\,II] -- BPT diagram to explore the gas properties. The [S\,II] emission lines are sensitive to the hardening of the ionization field, indicating that a non-thermal nuclear source might be producing the observed emission \citep{2006MNRAS.371.1559Groves}. 

The spatially resolved [S\,II] -- BPT diagrams reveal that most of the points are located in the Seyfert region for SDSS J030903+003846 (Fig.~\ref{fig:bpt_j030903}), but the [S\,II] emission is observed only in the inner $\sim$1 arcec radius.  For SDSS J033549-00391, the [S\,II] emission is observed over most of the GMOS FoV and the [S\,II] -- BPT diagram for this galaxy (Fig.~\ref{fig:bpt_j033549}) presents most of the points in the Seyfert/LINER region. Some points are observed in the the region occupied by star-forming galaxies, mostly from locations west of the nucleus, in the region where the emission line fluxes present their maximum values. The [S\,II] -- BPT diagram for  SDSS J033553-003946 present most points in the star-forming region, with some spaxels consistent with AGN ionization in regions farther than $\sim$1 arcsec from the nucleus.  A similar behaviour is observed for SDSS J225236-00331 (Fig.~\ref{fig:bpt_j225236}), but with the nuclear emission being consistent with AGN activity, and regions close to the gas emission peak consistent with gas ionization by young stars. Thus, our GMOS data reveal that both phenomena, AGN and star formation, are likely responsible for the observed gas emission in the four galaxies.

An AGN scenario can be supported by multi wavelength observations, with X-ray and radio emission being important to distinguish between stellar processes and nuclear activity, when optical observations are inconclusive \citep{2011Natur.470...66Reines, 2015ApJ...805...12Lemons, 2020ApJ...888...36Reines, 2020MNRAS.495L..71Yang, 2021ApJ...910....5Molina, 2024MNRAS.527.1962Bykov}. 
Radio and X-ray observations from the compact dwarf galaxy Henize 2-10 reveals a massive BH (M$_{BH} \sim 10^6$ M$_\odot$, \citep{2011Natur.470...66Reines}) with detections of an unresolved radio nuclear source  \citep{2011Natur.470...66Reines, 2012ApJ...750L..24Reines} and co-spatial with a compact hard X-ray source \citep{2011Natur.470...66Reines, 2016ApJ...830L..35Reines}. Additionally, near infrared integral field observations confirm the presence of a BH by identifying the ionization mechanism in the nuclear region where compact radio and X-ray emission were detected \citep{2020MNRAS.494.2004Riffel}. 
However, some studies propose that the X-ray emission observed for Henize 2-10 may not be associated with an AGN, but rather with another type of process, such as a supernova remnant \citep{2017A&A...604A.101Cresci, 2019MNRAS.485.5604Hebbar}. This example shows that the 
challenge of distinguishing between stellar processes and nuclear activity in the optical range extends to radio and X-ray wavelengths, especially at low Eddington accretion rates in dwarf galaxies \citep{2020ApJ...888...36Reines}. 
In radio wavelengths, synchrotron emission from AGN radio jets cannot be disguised by synchrotron emission from younger supernovae or supernovae remants, while in X-ray the emission can also be confused with TDE, X-ray binaries or with a background AGN. 
It also needs to be considered that even the continuum emitted by nuclear activity is not detected in radio or X-ray due to the low Eddington accretions rates, as mentioned by \cite{2022NatAs...6...26Reines}.
Identifying the presence of an accreting BH in dwarf galaxies is challenging, and therefore the demographics of these black holes may be underestimated due to the difficulty in detecting these objects.

Our results offer further evidence of the presence of AGN in the four galaxies, nonetheless there is no detection of radio emission to validate this scenario. Despite that, more comprehensive studies using deep multi-wavelength  observations (e.g. \cite{2023ApJ...959..116Nandi}) are imperative to distinguish between the contributions of AGN and star formation to the gas ionization.

\section{Conclusions} \label{sec:conclusions}

In this work, we used optical integral field spectroscopy of four four dwarf galaxies selected as being  candidates to host accreting massive black holes based on the detection of the \ironX\,  coronal emission line in single aperture spectra. We find that the four galaxies present extended gas emission covering the entire FoV (scales of $\sim$ 2.5 kpc $\times$ 1.5 kpc to $\sim$ 5.0 kpc $\times$ 3.8 kpc). The [S\,II] -- BPT diagnostic diagrams of the four galaxies present regions consistent with AGN photoionization. 
However, the \ironX\ emission line is not detected in any of the galaxies. We speculate that this non-detection  in our recent data can be due to a tidal disruption event that occurred in the past, when the SDSS data where observed, or a misidentification of this line, since the Si\,II\,$\lambda$6371 may contaminate this spectral region.  Ultimately, the gas kinematics across all galaxies aligns with the presence of perturbed rotating disks.

\section*{Acknowledgements}
Based on observations obtained at the Gemini Observatory, which is operated by the Association of Universities for Research in Astronomy, Inc., under a cooperative agreement with the NSF on behalf of the Gemini partnership: the National Science Foundation (United States), National Research Council (Canada), CONICYT (Chile), Ministerio de Ciencia, Tecnolog\'{i}a e Innovaci\'{o}n Productiva (Argentina), Minist\'{e}rio da Ci\^{e}ncia, Tecnologia e Inova\c{c}\~{a}o (Brazil), and Korea Astronomy and Space Science Institute (Republic of Korea). 
This research has made use of NASA's Astrophysics Data System Bibliographic Services. This research has made use of the NASA/IPAC Extragalactic Database (NED), which is operated by the Jet Propulsion Laboratory, California Institute of Technology, under contract with the National Aeronautics and Space Administration.

K.F.H. thank the financial support from Coordena\c c\~ao de Aperfei\c coamento de Pessoal de N\'ivel Superior - Brasil (CAPES) - Finance Code 001. 
R.A.R. acknowledges the support from Conselho Nacional de Desenvolvimento Cient\'ifico e Tecnol\'ogico (CNPq; Proj. 303450/2022-3, 403398/2023-1, \& 441722/2023-7), Funda\c c\~ao de Amparo \`a pesquisa do Estado do Rio Grande do Sul (FAPERGS; Proj. 21/2551-0002018-0), and CAPES (Proj. 88887.894973/2023-00). T.V.R. acknowledges the support from CNPq (Proj. 304584/2022-3).

\section*{Data Availability}

The data used in this work are publicly available online via the GEMINI archive https://archive.gemini.edu/searchform, under the program code GS-2022B-Q-320 and GN-2023A-Q-320. The processed data used in this paper will be shared on reasonable request to the corresponding author.



\bibliographystyle{mnras}
\bibliography{article_MNRAS} 

\begin{thebibliography}{}
\makeatletter
\relax
\def\mn@urlcharsother{\let\do\@makeother \do\$\do\&\do\#\do\^\do\_\do\%\do\~}
\def\mn@doi{\begingroup\mn@urlcharsother \@ifnextchar [ {\mn@doi@}
  {\mn@doi@[]}}
\def\mn@doi@[#1]#2{\def\@tempa{#1}\ifx\@tempa\@empty \href
  {http://dx.doi.org/#2} {doi:#2}\else \href {http://dx.doi.org/#2} {#1}\fi
  \endgroup}
\def\mn@eprint#1#2{\mn@eprint@#1:#2::\@nil}
\def\mn@eprint@arXiv#1{\href {http://arxiv.org/abs/#1} {{\tt arXiv:#1}}}
\def\mn@eprint@dblp#1{\href {http://dblp.uni-trier.de/rec/bibtex/#1.xml}
  {dblp:#1}}
\def\mn@eprint@#1:#2:#3:#4\@nil{\def\@tempa {#1}\def\@tempb {#2}\def\@tempc
  {#3}\ifx \@tempc \@empty \let \@tempc \@tempb \let \@tempb \@tempa \fi \ifx
  \@tempb \@empty \def\@tempb {arXiv}\fi \@ifundefined
  {mn@eprint@\@tempb}{\@tempb:\@tempc}{\expandafter \expandafter \csname
  mn@eprint@\@tempb\endcsname \expandafter{\@tempc}}}

\bibitem[\protect\citeauthoryear{{Allington-Smith} et~al.,}{{Allington-Smith}
  et~al.}{2002}]{AllingtonSmith_2002}
{Allington-Smith} J.,  et~al., 2002, \mn@doi [\pasp] {10.1086/341712}, \href
  {https://ui.adsabs.harvard.edu/abs/2002PASP..114..892A} {114, 892}

\bibitem[\protect\citeauthoryear{{Baldwin}, {Phillips}  \&
  {Terlevich}}{{Baldwin} et~al.}{1981}]{bpt_1981}
{Baldwin} J.~A.,  {Phillips} M.~M.,   {Terlevich} R.,  1981, \mn@doi [\pasp]
  {10.1086/130766}, \href
  {https://ui.adsabs.harvard.edu/abs/1981PASP...93....5B} {93, 5}

\bibitem[\protect\citeauthoryear{{Bellovary}, {Volonteri}, {Governato}, {Shen},
  {Quinn}  \& {Wadsley}}{{Bellovary}
  et~al.}{2011}]{2011ApJ...742...13Bellovary}
{Bellovary} J.,  {Volonteri} M.,  {Governato} F.,  {Shen} S.,  {Quinn} T.,
  {Wadsley} J.,  2011, \mn@doi [\apj] {10.1088/0004-637X/742/1/13}, \href
  {https://ui.adsabs.harvard.edu/abs/2011ApJ...742...13B} {742, 13}

\bibitem[\protect\citeauthoryear{{Bianchin} et~al.,}{{Bianchin}
  et~al.}{2022}]{Bianchin_22}
{Bianchin} M.,  et~al., 2022, \mn@doi [\mnras] {10.1093/mnras/stab3468}, \href
  {https://ui.adsabs.harvard.edu/abs/2022MNRAS.510..639B} {510, 639}

\bibitem[\protect\citeauthoryear{{Bykov}, {Gilfanov}  \& {Sunyaev}}{{Bykov}
  et~al.}{2024}]{2024MNRAS.527.1962Bykov}
{Bykov} S.~D.,  {Gilfanov} M.~R.,   {Sunyaev} R.~A.,  2024, \mn@doi [\mnras]
  {10.1093/mnras/stad3355}, \href
  {https://ui.adsabs.harvard.edu/abs/2024MNRAS.527.1962B} {527, 1962}

\bibitem[\protect\citeauthoryear{{Cann}, {Satyapal}, {Abel}, {Blecha},
  {Mushotzky}, {Reynolds}  \& {Secrest}}{{Cann}
  et~al.}{2019}]{2019ApJ...870L...2Cann}
{Cann} J.~M.,  {Satyapal} S.,  {Abel} N.~P.,  {Blecha} L.,  {Mushotzky} R.~F.,
  {Reynolds} C.~S.,   {Secrest} N.~J.,  2019, \mn@doi [\apjl]
  {10.3847/2041-8213/aaf88d}, \href
  {https://ui.adsabs.harvard.edu/abs/2019ApJ...870L...2C} {870, L2}

\bibitem[\protect\citeauthoryear{{Cattaneo} et~al.,}{{Cattaneo}
  et~al.}{2009}]{2009Natur.460..213Cattaneo}
{Cattaneo} A.,  et~al., 2009, \mn@doi [\nat] {10.1038/nature08135}, \href
  {https://ui.adsabs.harvard.edu/abs/2009Natur.460..213C} {460, 213}

\bibitem[\protect\citeauthoryear{{Cresci}, {Vanzi}, {Telles}, {Lanzuisi},
  {Brusa}, {Mingozzi}, {Sauvage}  \& {Johnson}}{{Cresci}
  et~al.}{2017}]{2017A&A...604A.101Cresci}
{Cresci} G.,  {Vanzi} L.,  {Telles} E.,  {Lanzuisi} G.,  {Brusa} M.,
  {Mingozzi} M.,  {Sauvage} M.,   {Johnson} K.,  2017, \mn@doi [\aap]
  {10.1051/0004-6361/201730876}, \href
  {https://ui.adsabs.harvard.edu/abs/2017A&A...604A.101C} {604, A101}

\bibitem[\protect\citeauthoryear{{Dashyan}, {Silk}, {Mamon}, {Dubois}  \&
  {Hartwig}}{{Dashyan} et~al.}{2018}]{2018MNRAS.473.5698Dashyan}
{Dashyan} G.,  {Silk} J.,  {Mamon} G.~A.,  {Dubois} Y.,   {Hartwig} T.,  2018,
  \mn@doi [\mnras] {10.1093/mnras/stx2716}, \href
  {https://ui.adsabs.harvard.edu/abs/2018MNRAS.473.5698D} {473, 5698}

\bibitem[\protect\citeauthoryear{{Di Matteo}, {Springel}  \& {Hernquist}}{{Di
  Matteo} et~al.}{2005}]{2005Natur.433..604DiMatteo}
{Di Matteo} T.,  {Springel} V.,   {Hernquist} L.,  2005, \mn@doi [\nat]
  {10.1038/nature03335}, \href
  {https://ui.adsabs.harvard.edu/abs/2005Natur.433..604D} {433, 604}

\bibitem[\protect\citeauthoryear{{Esquej}, {Saxton}, {Freyberg}, {Read},
  {Altieri}, {Sanchez-Portal}  \& {Hasinger}}{{Esquej}
  et~al.}{2007}]{2007A&A...462L..49Esquej}
{Esquej} P.,  {Saxton} R.~D.,  {Freyberg} M.~J.,  {Read} A.~M.,  {Altieri} B.,
  {Sanchez-Portal} M.,   {Hasinger} G.,  2007, \mn@doi [\aap]
  {10.1051/0004-6361:20066072}, \href
  {https://ui.adsabs.harvard.edu/abs/2007A&A...462L..49E} {462, L49}

\bibitem[\protect\citeauthoryear{{Fabian}}{{Fabian}}{2012}]{2012ARA&A..50..455Fabian}
{Fabian} A.~C.,  2012, \mn@doi [\araa] {10.1146/annurev-astro-081811-125521},
  \href {https://ui.adsabs.harvard.edu/abs/2012ARA&A..50..455F} {50, 455}

\bibitem[\protect\citeauthoryear{{Fischer} et~al.,}{{Fischer}
  et~al.}{2017}]{2017ApJ...834...30Fischer}
{Fischer} T.~C.,  et~al., 2017, \mn@doi [\apj] {10.3847/1538-4357/834/1/30},
  \href {https://ui.adsabs.harvard.edu/abs/2017ApJ...834...30F} {834, 30}

\bibitem[\protect\citeauthoryear{{Fonseca-Faria}, {Rodr{\'\i}guez-Ardila},
  {Contini}  \& {Reynaldi}}{{Fonseca-Faria}
  et~al.}{2021}]{2021MNRAS.506.3831Fonseca-Faria}
{Fonseca-Faria} M.~A.,  {Rodr{\'\i}guez-Ardila} A.,  {Contini} M.,   {Reynaldi}
  V.,  2021, \mn@doi [\mnras] {10.1093/mnras/stab1806}, \href
  {https://ui.adsabs.harvard.edu/abs/2021MNRAS.506.3831F} {506, 3831}

\bibitem[\protect\citeauthoryear{{F{\"o}rster Schreiber} et~al.,}{{F{\"o}rster
  Schreiber} et~al.}{2019}]{Forster_2019}
{F{\"o}rster Schreiber} N.~M.,  et~al., 2019, \mn@doi [\apj]
  {10.3847/1538-4357/ab0ca2}, \href
  {https://ui.adsabs.harvard.edu/abs/2019ApJ...875...21F} {875, 21}

\bibitem[\protect\citeauthoryear{{French}, {Wevers}, {Law-Smith}, {Graur}  \&
  {Zabludoff}}{{French} et~al.}{2020}]{2020SSRv..216...32French}
{French} K.~D.,  {Wevers} T.,  {Law-Smith} J.,  {Graur} O.,   {Zabludoff}
  A.~I.,  2020, \mn@doi [\ssr] {10.1007/s11214-020-00657-y}, \href
  {https://ui.adsabs.harvard.edu/abs/2020SSRv..216...32F} {216, 32}

\bibitem[\protect\citeauthoryear{{Gezari}}{{Gezari}}{2021}]{2021ARA&A..59...21Gezari}
{Gezari} S.,  2021, \mn@doi [\araa] {10.1146/annurev-astro-111720-030029},
  \href {https://ui.adsabs.harvard.edu/abs/2021ARA&A..59...21G} {59, 21}

\bibitem[\protect\citeauthoryear{Gimeno et~al.,}{Gimeno
  et~al.}{2016}]{2016SPIE.9908E..2SGimeno}
Gimeno G.,  et~al., 2016, in Evans C.~J.,  Simard L.,   Takami H.,  eds,  Vol.
  9908, Ground-based and Airborne Instrumentation for Astronomy VI. SPIE, p.
  99082S, \mn@doi{10.1117/12.2233883}

\bibitem[\protect\citeauthoryear{{Greene}, {Strader}  \& {Ho}}{{Greene}
  et~al.}{2020}]{2020ARA&A..58..257Greene}
{Greene} J.~E.,  {Strader} J.,   {Ho} L.~C.,  2020, \mn@doi [\araa]
  {10.1146/annurev-astro-032620-021835}, \href
  {https://ui.adsabs.harvard.edu/abs/2020ARA&A..58..257G} {58, 257}

\bibitem[\protect\citeauthoryear{{Groves}, {Heckman}  \& {Kauffmann}}{{Groves}
  et~al.}{2006}]{2006MNRAS.371.1559Groves}
{Groves} B.~A.,  {Heckman} T.~M.,   {Kauffmann} G.,  2006, \mn@doi [\mnras]
  {10.1111/j.1365-2966.2006.10812.x}, \href
  {https://ui.adsabs.harvard.edu/abs/2006MNRAS.371.1559G} {371, 1559}

\bibitem[\protect\citeauthoryear{{Harrison}}{{Harrison}}{2017}]{2017NatAs...1E.165Harrison}
{Harrison} C.~M.,  2017, \mn@doi [Nature Astronomy] {10.1038/s41550-017-0165},
  \href {https://ui.adsabs.harvard.edu/abs/2017NatAs...1E.165H} {1, 0165}

\bibitem[\protect\citeauthoryear{{Hebbar}, {Heinke}, {Sivakoff}  \&
  {Shaw}}{{Hebbar} et~al.}{2019}]{2019MNRAS.485.5604Hebbar}
{Hebbar} P.~R.,  {Heinke} C.~O.,  {Sivakoff} G.~R.,   {Shaw} A.~W.,  2019,
  \mn@doi [\mnras] {10.1093/mnras/stz553}, \href
  {https://ui.adsabs.harvard.edu/abs/2019MNRAS.485.5604H} {485, 5604}

\bibitem[\protect\citeauthoryear{{Herenz}, {Micheva}, {Weilbacher},
  {Monreal-Ibero}, {Hayes}, {Anders}  \& {Rivinius}}{{Herenz}
  et~al.}{2023}]{2023RNAAS...7...99Herenz}
{Herenz} E.~C.,  {Micheva} G.,  {Weilbacher} P.~M.,  {Monreal-Ibero} A.,
  {Hayes} M.,  {Anders} F.,   {Rivinius} T.,  2023, \mn@doi [Research Notes of
  the American Astronomical Society] {10.3847/2515-5172/acd69e}, \href
  {https://ui.adsabs.harvard.edu/abs/2023RNAAS...7...99H} {7, 99}

\bibitem[\protect\citeauthoryear{{Hills}}{{Hills}}{1975}]{1975Natur.254..295Hills}
{Hills} J.~G.,  1975, \mn@doi [\nat] {10.1038/254295a0}, \href
  {https://ui.adsabs.harvard.edu/abs/1975Natur.254..295H} {254, 295}

\bibitem[\protect\citeauthoryear{{Ho}}{{Ho}}{2008}]{2008ARA&A..46..475Ho}
{Ho} L.~C.,  2008, \mn@doi [\araa] {10.1146/annurev.astro.45.051806.110546},
  \href {https://ui.adsabs.harvard.edu/abs/2008ARA&A..46..475H} {46, 475}

\bibitem[\protect\citeauthoryear{{Hopkins} \& {Hernquist}}{{Hopkins} \&
  {Hernquist}}{2006}]{2006ApJS..166....1Hopkins}
{Hopkins} P.~F.,  {Hernquist} L.,  2006, \mn@doi [\apjs] {10.1086/505753},
  \href {https://ui.adsabs.harvard.edu/abs/2006ApJS..166....1H} {166, 1}

\bibitem[\protect\citeauthoryear{{Jeon}, {Pawlik}, {Greif}, {Glover}, {Bromm},
  {Milosavljevi{\'c}}  \& {Klessen}}{{Jeon}
  et~al.}{2012}]{2012ApJ...754...34Jeon}
{Jeon} M.,  {Pawlik} A.~H.,  {Greif} T.~H.,  {Glover} S. C.~O.,  {Bromm} V.,
  {Milosavljevi{\'c}} M.,   {Klessen} R.~S.,  2012, \mn@doi [\apj]
  {10.1088/0004-637X/754/1/34}, \href
  {https://ui.adsabs.harvard.edu/abs/2012ApJ...754...34J} {754, 34}

\bibitem[\protect\citeauthoryear{{Juneau} et~al.,}{{Juneau}
  et~al.}{2022}]{Juneau_22}
{Juneau} S.,  et~al., 2022, \mn@doi [\apj] {10.3847/1538-4357/ac425f}, \href
  {https://ui.adsabs.harvard.edu/abs/2022ApJ...925..203J} {925, 203}

\bibitem[\protect\citeauthoryear{{Kewley}, {Groves}, {Kauffmann}  \&
  {Heckman}}{{Kewley} et~al.}{2006}]{2006MNRAS.372..961Kewley}
{Kewley} L.~J.,  {Groves} B.,  {Kauffmann} G.,   {Heckman} T.,  2006, \mn@doi
  [\mnras] {10.1111/j.1365-2966.2006.10859.x}, \href
  {https://ui.adsabs.harvard.edu/abs/2006MNRAS.372..961K} {372, 961}

\bibitem[\protect\citeauthoryear{{Kewley}, {Dopita}, {Leitherer}, {Dav{\'e}},
  {Yuan}, {Allen}, {Groves}  \& {Sutherland}}{{Kewley}
  et~al.}{2013}]{2013ApJ...774..100Kewley}
{Kewley} L.~J.,  {Dopita} M.~A.,  {Leitherer} C.,  {Dav{\'e}} R.,  {Yuan} T.,
  {Allen} M.,  {Groves} B.,   {Sutherland} R.,  2013, \mn@doi [\apj]
  {10.1088/0004-637X/774/2/100}, \href
  {https://ui.adsabs.harvard.edu/abs/2013ApJ...774..100K} {774, 100}

\bibitem[\protect\citeauthoryear{{Kimbrell}, {Reines}, {Greene}  \&
  {Geha}}{{Kimbrell} et~al.}{2023}]{Kimbrell23}
{Kimbrell} S.~J.,  {Reines} A.~E.,  {Greene} J.~E.,   {Geha} M.,  2023, \mn@doi
  [\apj] {10.3847/1538-4357/acf762}, \href
  {https://ui.adsabs.harvard.edu/abs/2023ApJ...958..115K} {958, 115}

\bibitem[\protect\citeauthoryear{{Kormendy} \& {Ho}}{{Kormendy} \&
  {Ho}}{2013}]{2013ARA&A..51..511Kormendy}
{Kormendy} J.,  {Ho} L.~C.,  2013, \mn@doi [\araa]
  {10.1146/annurev-astro-082708-101811}, \href
  {https://ui.adsabs.harvard.edu/abs/2013ARA&A..51..511K} {51, 511}

\bibitem[\protect\citeauthoryear{{Kormendy} \& {Richstone}}{{Kormendy} \&
  {Richstone}}{1995}]{1995ARA&A..33..581Kormendy}
{Kormendy} J.,  {Richstone} D.,  1995, \mn@doi [\araa]
  {10.1146/annurev.aa.33.090195.003053}, \href
  {https://ui.adsabs.harvard.edu/abs/1995ARA&A..33..581K} {33, 581}

\bibitem[\protect\citeauthoryear{Kramida, Ralchenko, {Reader}  \& {NIST ASD
  Team}}{Kramida et~al.}{2022}]{kramida2022_NIST}
Kramida A.,  Ralchenko Y.,  {Reader} J.,   {NIST ASD Team} 2022, {NIST Atomic
  Spectra Database (version 5.10)}, \mn@doi{https://doi.org/10.18434/T4W30F},
  \url {https://physics.nist.gov/asd}

\bibitem[\protect\citeauthoryear{{Lemons}, {Reines}, {Plotkin}, {Gallo}  \&
  {Greene}}{{Lemons} et~al.}{2015}]{2015ApJ...805...12Lemons}
{Lemons} S.~M.,  {Reines} A.~E.,  {Plotkin} R.~M.,  {Gallo} E.,   {Greene}
  J.~E.,  2015, \mn@doi [\apj] {10.1088/0004-637X/805/1/12}, \href
  {https://ui.adsabs.harvard.edu/abs/2015ApJ...805...12L} {805, 12}

\bibitem[\protect\citeauthoryear{{Lin} et~al.,}{{Lin}
  et~al.}{2018}]{2018NatAs...2..656Lin}
{Lin} D.,  et~al., 2018, \mn@doi [Nature Astronomy]
  {10.1038/s41550-018-0493-1}, \href
  {https://ui.adsabs.harvard.edu/abs/2018NatAs...2..656L} {2, 656}

\bibitem[\protect\citeauthoryear{{Lutz} et~al.,}{{Lutz} et~al.}{2020}]{Lutz_20}
{Lutz} D.,  et~al., 2020, \mn@doi [\aap] {10.1051/0004-6361/201936803}, \href
  {https://ui.adsabs.harvard.edu/abs/2020A&A...633A.134L} {633, A134}

\bibitem[\protect\citeauthoryear{{Magorrian} \& {Tremaine}}{{Magorrian} \&
  {Tremaine}}{1999}]{1999MNRAS.309..447Magorrian}
{Magorrian} J.,  {Tremaine} S.,  1999, \mn@doi [\mnras]
  {10.1046/j.1365-8711.1999.02853.x}, \href
  {https://ui.adsabs.harvard.edu/abs/1999MNRAS.309..447M} {309, 447}

\bibitem[\protect\citeauthoryear{{Magorrian} et~al.,}{{Magorrian}
  et~al.}{1998}]{1998AJ....115.2285Magorrian}
{Magorrian} J.,  et~al., 1998, \mn@doi [\aj] {10.1086/300353}, \href
  {https://ui.adsabs.harvard.edu/abs/1998AJ....115.2285M} {115, 2285}

\bibitem[\protect\citeauthoryear{{Maksym}, {Lin}  \& {Irwin}}{{Maksym}
  et~al.}{2014}]{2014ApJ...792L..29Maksym}
{Maksym} W.~P.,  {Lin} D.,   {Irwin} J.~A.,  2014, \mn@doi [\apjl]
  {10.1088/2041-8205/792/2/L29}, \href
  {https://ui.adsabs.harvard.edu/abs/2014ApJ...792L..29M} {792, L29}

\bibitem[\protect\citeauthoryear{{Manzano-King}, {Canalizo}  \&
  {Sales}}{{Manzano-King} et~al.}{2019}]{2019ApJ...884...54ManzanoKing}
{Manzano-King} C.~M.,  {Canalizo} G.,   {Sales} L.~V.,  2019, \mn@doi [\apj]
  {10.3847/1538-4357/ab4197}, \href
  {https://ui.adsabs.harvard.edu/abs/2019ApJ...884...54M} {884, 54}

\bibitem[\protect\citeauthoryear{{Menezes}, {Ricci}, {Steiner}, {da Silva},
  {Ferrari}  \& {Borges}}{{Menezes} et~al.}{2019}]{menezes2019}
{Menezes} R.~B.,  {Ricci} T.~V.,  {Steiner} J.~E.,  {da Silva} P.,  {Ferrari}
  F.,   {Borges} B.~W.,  2019, \mn@doi [\mnras] {10.1093/mnras/sty3337}, \href
  {http://adsabs.harvard.edu/abs/2019MNRAS.483.3700M} {483, 3700}

\bibitem[\protect\citeauthoryear{{Mezcua} \& {Dom{\'\i}nguez
  S{\'a}nchez}}{{Mezcua} \& {Dom{\'\i}nguez S{\'a}nchez}}{2020}]{Mezcua20}
{Mezcua} M.,  {Dom{\'\i}nguez S{\'a}nchez} H.,  2020, \mn@doi [\apjl]
  {10.3847/2041-8213/aba199}, \href
  {https://ui.adsabs.harvard.edu/abs/2020ApJ...898L..30M} {898, L30}

\bibitem[\protect\citeauthoryear{{Mezcua} \& {Dom{\'\i}nguez
  S{\'a}nchez}}{{Mezcua} \& {Dom{\'\i}nguez S{\'a}nchez}}{2024}]{Mezcua24}
{Mezcua} M.,  {Dom{\'\i}nguez S{\'a}nchez} H.,  2024, \mn@doi [\mnras]
  {10.1093/mnras/stae292}, \href
  {https://ui.adsabs.harvard.edu/abs/2024MNRAS.528.5252M} {528, 5252}

\bibitem[\protect\citeauthoryear{{Molina}, {Reines}, {Greene}, {Darling}  \&
  {Condon}}{{Molina} et~al.}{2021a}]{2021ApJ...910....5Molina}
{Molina} M.,  {Reines} A.~E.,  {Greene} J.~E.,  {Darling} J.,   {Condon} J.~J.,
   2021a, \mn@doi [\apj] {10.3847/1538-4357/abe120}, \href
  {https://ui.adsabs.harvard.edu/abs/2021ApJ...910....5M} {910, 5}

\bibitem[\protect\citeauthoryear{{Molina}, {Reines}, {Latimer}, {Baldassare}
  \& {Salehirad}}{{Molina} et~al.}{2021b}]{2021ApJ...922..155Molina_sample}
{Molina} M.,  {Reines} A.~E.,  {Latimer} L.~J.,  {Baldassare} V.,   {Salehirad}
  S.,  2021b, \mn@doi [\apj] {10.3847/1538-4357/ac1ffa}, \href
  {https://ui.adsabs.harvard.edu/abs/2021ApJ...922..155M} {922, 155}

\bibitem[\protect\citeauthoryear{{Nandi} et~al.,}{{Nandi}
  et~al.}{2023}]{2023ApJ...959..116Nandi}
{Nandi} P.,  et~al., 2023, \mn@doi [\apj] {10.3847/1538-4357/ad0c57}, \href
  {https://ui.adsabs.harvard.edu/abs/2023ApJ...959..116N} {959, 116}

\bibitem[\protect\citeauthoryear{{Osterbrock} \& {Ferland}}{{Osterbrock} \&
  {Ferland}}{2006}]{Osterbrock2006}
{Osterbrock} D.~E.,  {Ferland} G.~J.,  2006, {Astrophysics of gaseous nebulae
  and active galactic nuclei}.
Sausalito, CA: University Science Books

\bibitem[\protect\citeauthoryear{{Penny} et~al.,}{{Penny}
  et~al.}{2018}]{Penny_18}
{Penny} S.~J.,  et~al., 2018, \mn@doi [\mnras] {10.1093/mnras/sty202}, \href
  {https://ui.adsabs.harvard.edu/abs/2018MNRAS.476..979P} {476, 979}

\bibitem[\protect\citeauthoryear{{Reefe} et~al.,}{{Reefe}
  et~al.}{2023}]{2023ApJ...946L..38Reefe}
{Reefe} M.,  et~al., 2023, \mn@doi [\apjl] {10.3847/2041-8213/acb4e4}, \href
  {https://ui.adsabs.harvard.edu/abs/2023ApJ...946L..38R} {946, L38}

\bibitem[\protect\citeauthoryear{{Reines}}{{Reines}}{2022}]{2022NatAs...6...26Reines}
{Reines} A.~E.,  2022, \mn@doi [Nature Astronomy] {10.1038/s41550-021-01556-0},
  \href {https://ui.adsabs.harvard.edu/abs/2022NatAs...6...26R} {6, 26}

\bibitem[\protect\citeauthoryear{{Reines} \& {Deller}}{{Reines} \&
  {Deller}}{2012}]{2012ApJ...750L..24Reines}
{Reines} A.~E.,  {Deller} A.~T.,  2012, \mn@doi [\apjl]
  {10.1088/2041-8205/750/1/L24}, \href
  {https://ui.adsabs.harvard.edu/abs/2012ApJ...750L..24R} {750, L24}

\bibitem[\protect\citeauthoryear{{Reines}, {Sivakoff}, {Johnson}  \&
  {Brogan}}{{Reines} et~al.}{2011}]{2011Natur.470...66Reines}
{Reines} A.~E.,  {Sivakoff} G.~R.,  {Johnson} K.~E.,   {Brogan} C.~L.,  2011,
  \mn@doi [\nat] {10.1038/nature09724}, \href
  {https://ui.adsabs.harvard.edu/abs/2011Natur.470...66R} {470, 66}

\bibitem[\protect\citeauthoryear{{Reines}, {Greene}  \& {Geha}}{{Reines}
  et~al.}{2013}]{2013ApJ...775..116Reines}
{Reines} A.~E.,  {Greene} J.~E.,   {Geha} M.,  2013, \mn@doi [\apj]
  {10.1088/0004-637X/775/2/116}, \href
  {https://ui.adsabs.harvard.edu/abs/2013ApJ...775..116R} {775, 116}

\bibitem[\protect\citeauthoryear{{Reines}, {Reynolds}, {Miller}, {Sivakoff},
  {Greene}, {Hickox}  \& {Johnson}}{{Reines}
  et~al.}{2016}]{2016ApJ...830L..35Reines}
{Reines} A.~E.,  {Reynolds} M.~T.,  {Miller} J.~M.,  {Sivakoff} G.~R.,
  {Greene} J.~E.,  {Hickox} R.~C.,   {Johnson} K.~E.,  2016, \mn@doi [\apjl]
  {10.3847/2041-8205/830/2/L35}, \href
  {https://ui.adsabs.harvard.edu/abs/2016ApJ...830L..35R} {830, L35}

\bibitem[\protect\citeauthoryear{{Reines}, {Condon}, {Darling}  \&
  {Greene}}{{Reines} et~al.}{2020}]{2020ApJ...888...36Reines}
{Reines} A.~E.,  {Condon} J.~J.,  {Darling} J.,   {Greene} J.~E.,  2020,
  \mn@doi [\apj] {10.3847/1538-4357/ab4999}, \href
  {https://ui.adsabs.harvard.edu/abs/2020ApJ...888...36R} {888, 36}

\bibitem[\protect\citeauthoryear{{Ricci}, {Steiner}  \& {Menezes}}{{Ricci}
  et~al.}{2014}]{Ricci2014a}
{Ricci} T.~V.,  {Steiner} J.~E.,   {Menezes} R.~B.,  2014, \mn@doi [\mnras]
  {10.1093/mnras/stu441}, \href
  {http://adsabs.harvard.edu/abs/2014MNRAS.440.2419R} {440, 2419}

\bibitem[\protect\citeauthoryear{{Riffel}}{{Riffel}}{2020}]{2020MNRAS.494.2004Riffel}
{Riffel} R.~A.,  2020, \mn@doi [\mnras] {10.1093/mnras/staa903}, \href
  {https://ui.adsabs.harvard.edu/abs/2020MNRAS.494.2004R} {494, 2004}

\bibitem[\protect\citeauthoryear{{Riffel} et~al.,}{{Riffel}
  et~al.}{2023}]{Rogemar_2023_kin}
{Riffel} R.~A.,  et~al., 2023, \mn@doi [\mnras] {10.1093/mnras/stad599}, \href
  {https://ui.adsabs.harvard.edu/abs/2023MNRAS.521.1832R} {521, 1832}

\bibitem[\protect\citeauthoryear{Ruschel-Dutra \& de Oliveira}{Ruschel-Dutra \&
  de~Oliveira}{2020}]{daniel_ruschel_dutra_2020_3945237}
Ruschel-Dutra D.,  de Oliveira B.~D.,  2020, danielrd6/ifscube v1.0,
  \mn@doi{10.5281/zenodo.3945237}

\bibitem[\protect\citeauthoryear{{Ruschel-Dutra} et~al.,}{{Ruschel-Dutra}
  et~al.}{2021}]{Ruschel-dutra_2021}
{Ruschel-Dutra} D.,  et~al., 2021, \mn@doi [\mnras] {10.1093/mnras/stab2058},
  \href {https://ui.adsabs.harvard.edu/abs/2021MNRAS.507...74R} {507, 74}

\bibitem[\protect\citeauthoryear{{Schutte} \& {Reines}}{{Schutte} \&
  {Reines}}{2022}]{2022Natur.601..329Schutte}
{Schutte} Z.,  {Reines} A.~E.,  2022, \mn@doi [\nat]
  {10.1038/s41586-021-04215-6}, \href
  {https://ui.adsabs.harvard.edu/abs/2022Natur.601..329S} {601, 329}

\bibitem[\protect\citeauthoryear{{Short} et~al.,}{{Short}
  et~al.}{2023}]{2023MNRAS.525.1568Short}
{Short} P.,  et~al., 2023, \mn@doi [\mnras] {10.1093/mnras/stad2270}, \href
  {https://ui.adsabs.harvard.edu/abs/2023MNRAS.525.1568S} {525, 1568}

\bibitem[\protect\citeauthoryear{{Silk} \& {Mamon}}{{Silk} \&
  {Mamon}}{2012}]{2012RAA....12..917Silk}
{Silk} J.,  {Mamon} G.~A.,  2012, \mn@doi [Research in Astronomy and
  Astrophysics] {10.1088/1674-4527/12/8/004}, \href
  {https://ui.adsabs.harvard.edu/abs/2012RAA....12..917S} {12, 917}

\bibitem[\protect\citeauthoryear{{Stone} \& {Metzger}}{{Stone} \&
  {Metzger}}{2016}]{2016MNRAS.455..859Stone}
{Stone} N.~C.,  {Metzger} B.~D.,  2016, \mn@doi [\mnras]
  {10.1093/mnras/stv2281}, \href
  {https://ui.adsabs.harvard.edu/abs/2016MNRAS.455..859S} {455, 859}

\bibitem[\protect\citeauthoryear{{Tremonti} et~al.,}{{Tremonti}
  et~al.}{2004}]{2004ApJ...613..898Tremonti}
{Tremonti} C.~A.,  et~al., 2004, \mn@doi [\apj] {10.1086/423264}, \href
  {https://ui.adsabs.harvard.edu/abs/2004ApJ...613..898T} {613, 898}

\bibitem[\protect\citeauthoryear{{U} et~al.,}{{U} et~al.}{2022}]{U_2022}
{U} V.,  et~al., 2022, \mn@doi [\apjl] {10.3847/2041-8213/ac961c}, \href
  {https://ui.adsabs.harvard.edu/abs/2022ApJ...940L...5U} {940, L5}

\bibitem[\protect\citeauthoryear{{Van Dokkum}}{{Van
  Dokkum}}{2001}]{van_Dokkum_2001}
{Van Dokkum} P.~G.,  2001, \mn@doi [\pasp] {10.1086/323894}, \href
  {https://ui.adsabs.harvard.edu/abs/2001PASP..113.1420V} {113, 1420}

\bibitem[\protect\citeauthoryear{{Wang} \& {Merritt}}{{Wang} \&
  {Merritt}}{2004}]{2004ApJ...600..149Wang}
{Wang} J.,  {Merritt} D.,  2004, \mn@doi [\apj] {10.1086/379767}, \href
  {https://ui.adsabs.harvard.edu/abs/2004ApJ...600..149W} {600, 149}

\bibitem[\protect\citeauthoryear{{Wang}, {Zhou}, {Komossa}, {Wang}, {Yuan}  \&
  {Yang}}{{Wang} et~al.}{2012}]{2012ApJ...749..115Wang}
{Wang} T.-G.,  {Zhou} H.-Y.,  {Komossa} S.,  {Wang} H.-Y.,  {Yuan} W.,   {Yang}
  C.,  2012, \mn@doi [\apj] {10.1088/0004-637X/749/2/115}, \href
  {https://ui.adsabs.harvard.edu/abs/2012ApJ...749..115W} {749, 115}

\bibitem[\protect\citeauthoryear{{Yang}, {Gurvits}, {Paragi}, {Frey}, {Conway},
  {Liu}  \& {Cui}}{{Yang} et~al.}{2020}]{2020MNRAS.495L..71Yang}
{Yang} J.,  {Gurvits} L.~I.,  {Paragi} Z.,  {Frey} S.,  {Conway} J.~E.,  {Liu}
  X.,   {Cui} L.,  2020, \mn@doi [\mnras] {10.1093/mnrasl/slaa052}, \href
  {https://ui.adsabs.harvard.edu/abs/2020MNRAS.495L..71Y} {495, L71}

\bibitem[\protect\citeauthoryear{{van Velzen}}{{van
  Velzen}}{2018}]{2018ApJ...852...72VanVelzen}
{van Velzen} S.,  2018, \mn@doi [\apj] {10.3847/1538-4357/aa998e}, \href
  {https://ui.adsabs.harvard.edu/abs/2018ApJ...852...72V} {852, 72}

\bibitem[\protect\citeauthoryear{{van Wassenhove}, {Volonteri}, {Walker}  \&
  {Gair}}{{van Wassenhove} et~al.}{2010}]{2010MNRAS.408.1139VanWassenhove}
{van Wassenhove} S.,  {Volonteri} M.,  {Walker} M.~G.,   {Gair} J.~R.,  2010,
  \mn@doi [\mnras] {10.1111/j.1365-2966.2010.17189.x}, \href
  {https://ui.adsabs.harvard.edu/abs/2010MNRAS.408.1139V} {408, 1139}

\makeatother
\end{thebibliography}




\appendix

\section{Emission line flux distributions and kinematics}\label{appendix}

\begin{figure*}
    \centering
    \includegraphics[width=\linewidth, trim=0cm 29cm 0cm 0cm, clip]{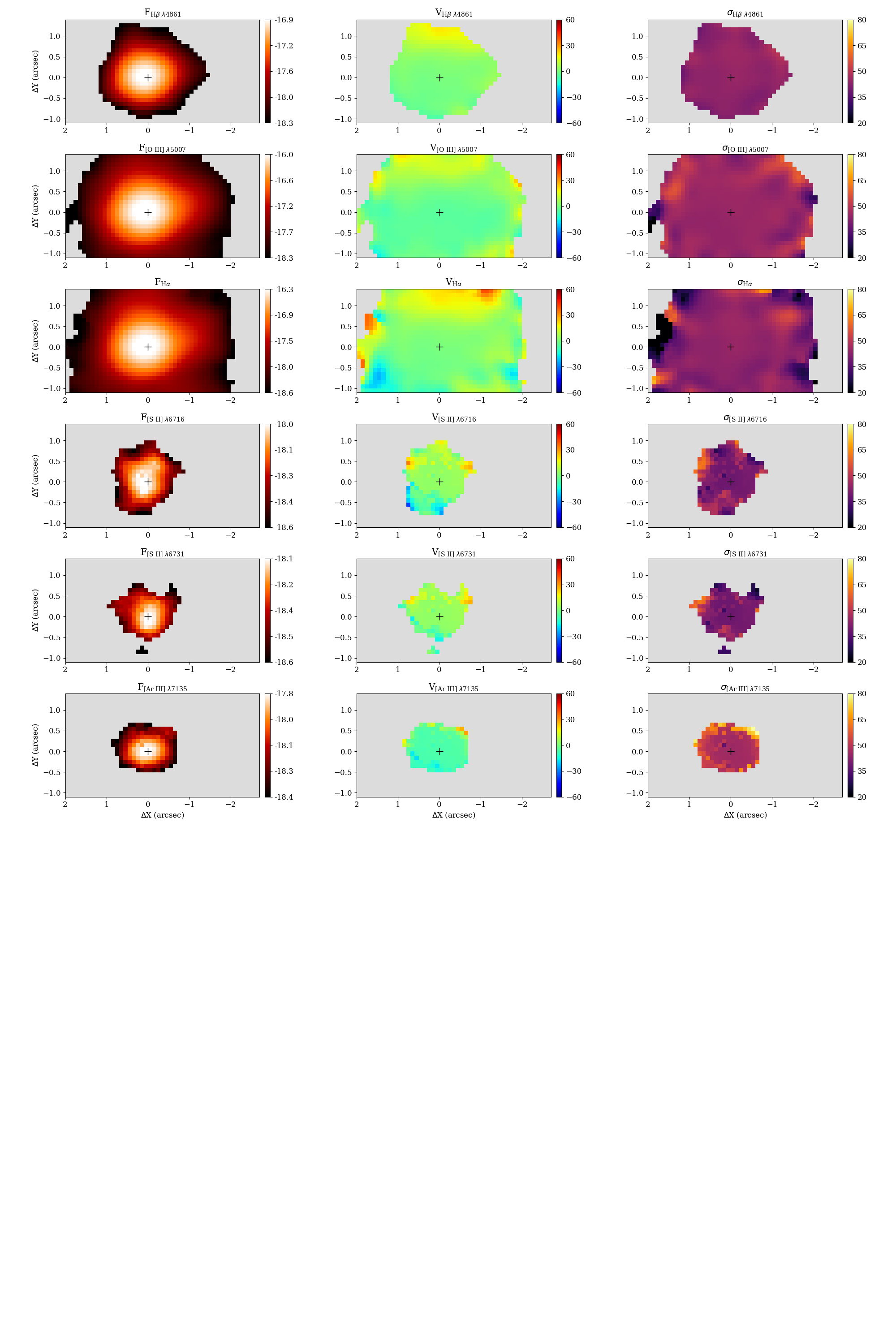}
    \caption{Emission line flux distribution and kinematics of the SDSS J030903+003846 galaxy. First panel: flux distribution in  erg\,s$^{-1}$\,cm$^{-2}$\,\AA$^{-1}$ and in logarithmic scale. Middle panel: observed radial velocity, in \kms. Third panel: Observed velocity dispersion, in \kms.}
    \label{fig:1g_j030903}
\end{figure*}

\begin{figure*}
    \centering
    \includegraphics[width=\linewidth, trim=0cm 22cm 0cm 0cm, clip]{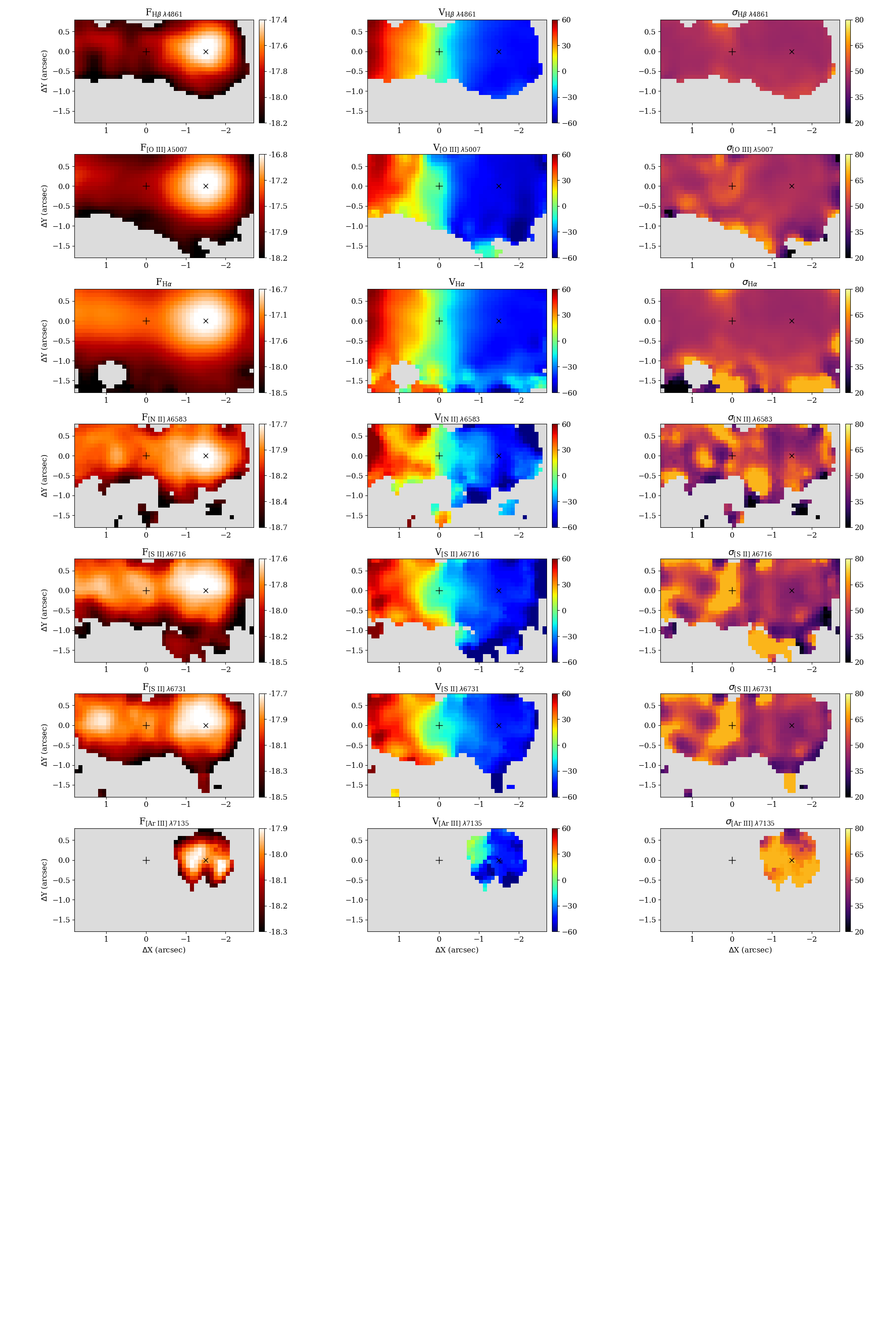}
    \caption{Same as Fig. \ref{fig:1g_j030903}, but for SDSS J033549-00391.}
    \label{fig:1g_j033549}
\end{figure*}

\begin{figure*}
    \centering
    \includegraphics[width=\linewidth, trim=0cm 22cm 0cm 0cm, clip]{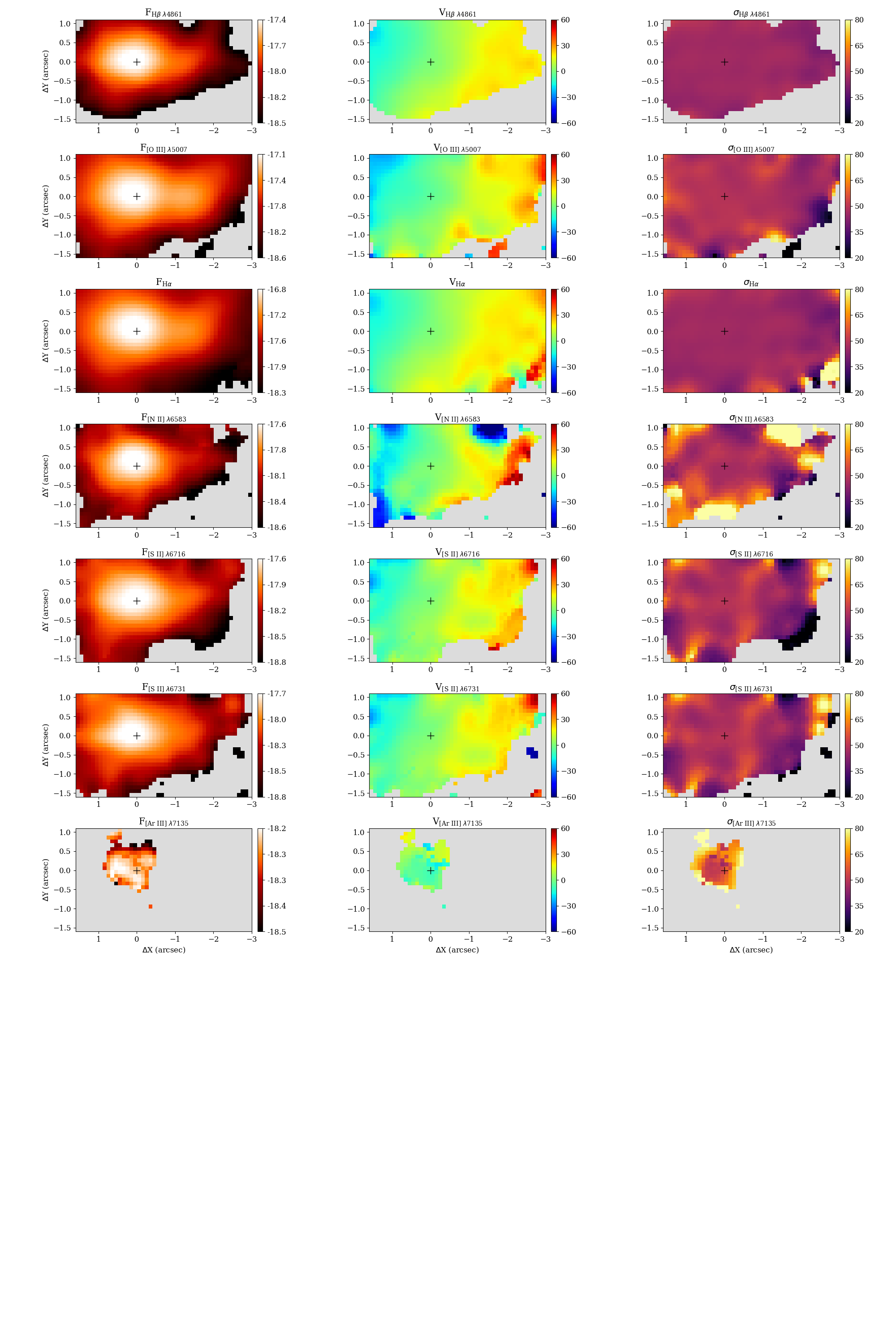}
    \caption{Same as Fig. \ref{fig:1g_j030903}, but for SDSS J033553-003946.}
    \label{fig:1g_j033553}
\end{figure*}

\begin{figure*}
    \centering
    \includegraphics[width=\linewidth, trim=0cm 6.5cm 0cm 0cm, clip]{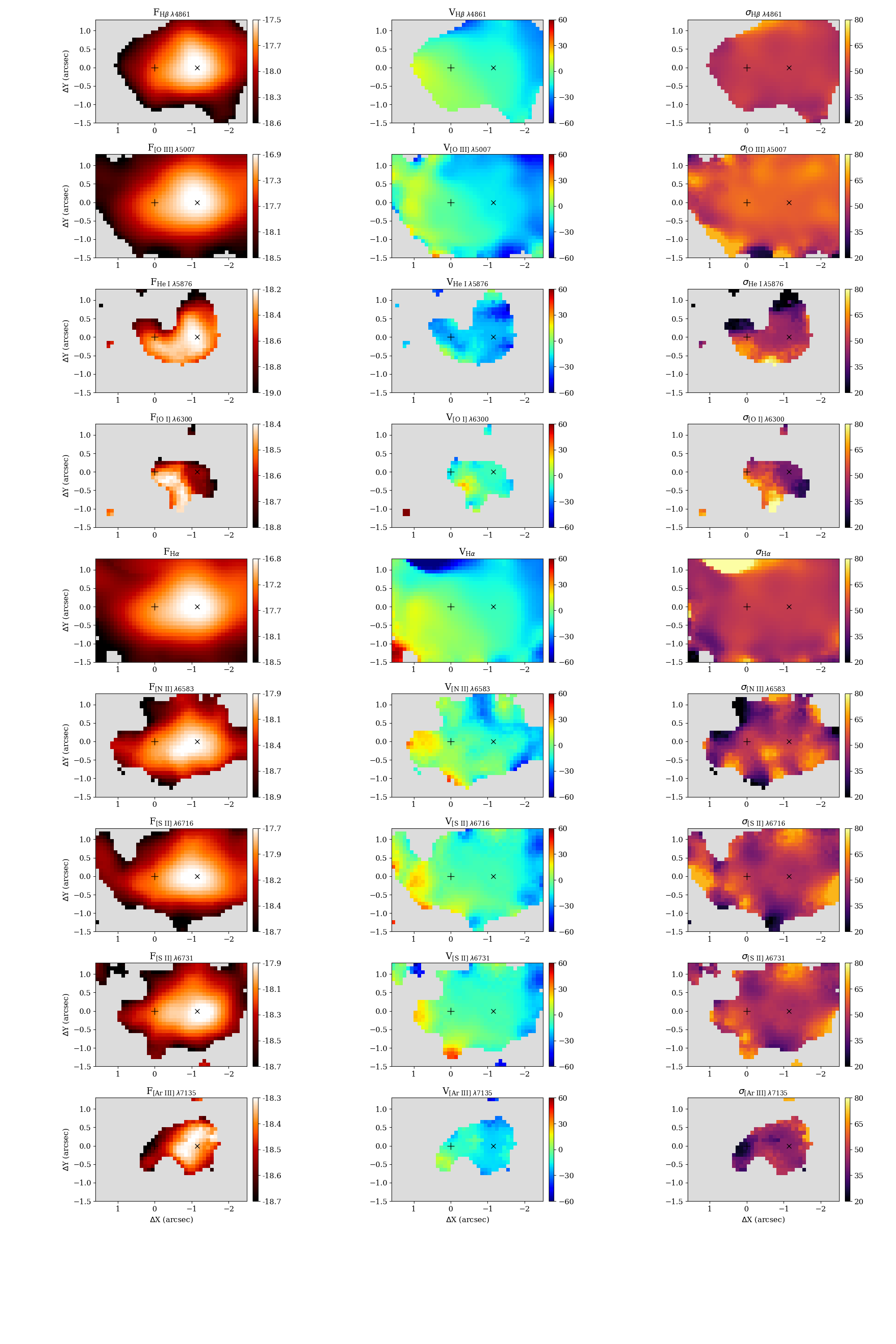}
    \caption{Same as Fig. \ref{fig:1g_j030903}, but for SDSS J225236-00331. The black cross mark the H$\alpha$ peak emission and the black plus sign represent the galaxy centre.}
    \label{fig:1g_j225236}
\end{figure*}

Figures~\ref{fig:1g_j030903} -- \ref{fig:1g_j225236} present the flux distributions and kinematic maps for all the emission lines detected for each galaxy.

\section{Spatially resolved diagnostic diagrams}

\begin{figure*}
    \centering
    \includegraphics[scale=0.4, trim=0cm 0cm 0cm 12cm, clip]{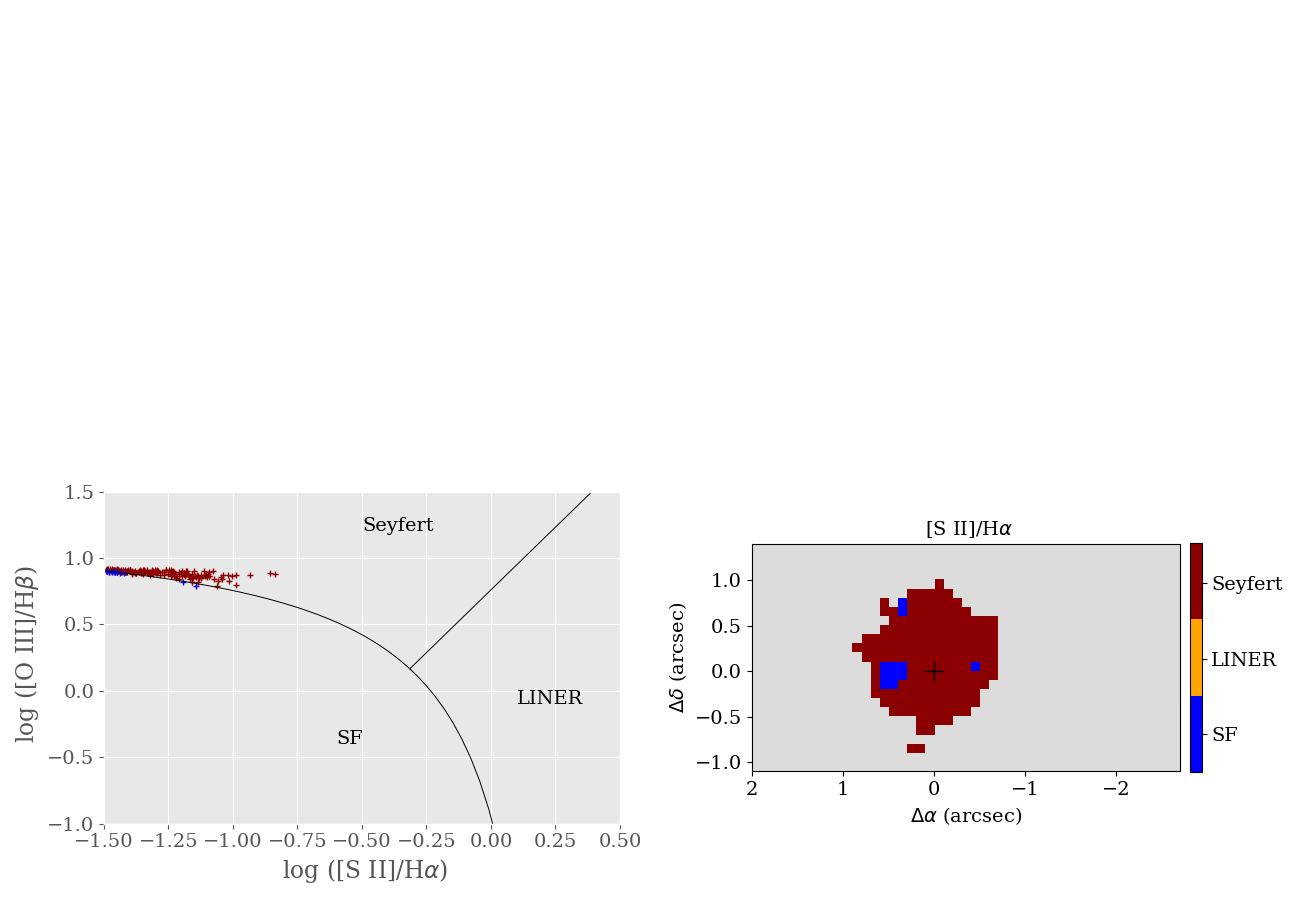}
    \caption{Optical BPT diagnostic diagram \citep{bpt_1981} for the SDSS J030903+003846 galaxy. Left panel: BPT -- \textbf{[S\,II]} diagram. Right panel: Spatially resolved excitation map.}
    \label{fig:bpt_j030903}
\end{figure*}

\begin{figure*}
    \centering
    \includegraphics[scale=0.4]{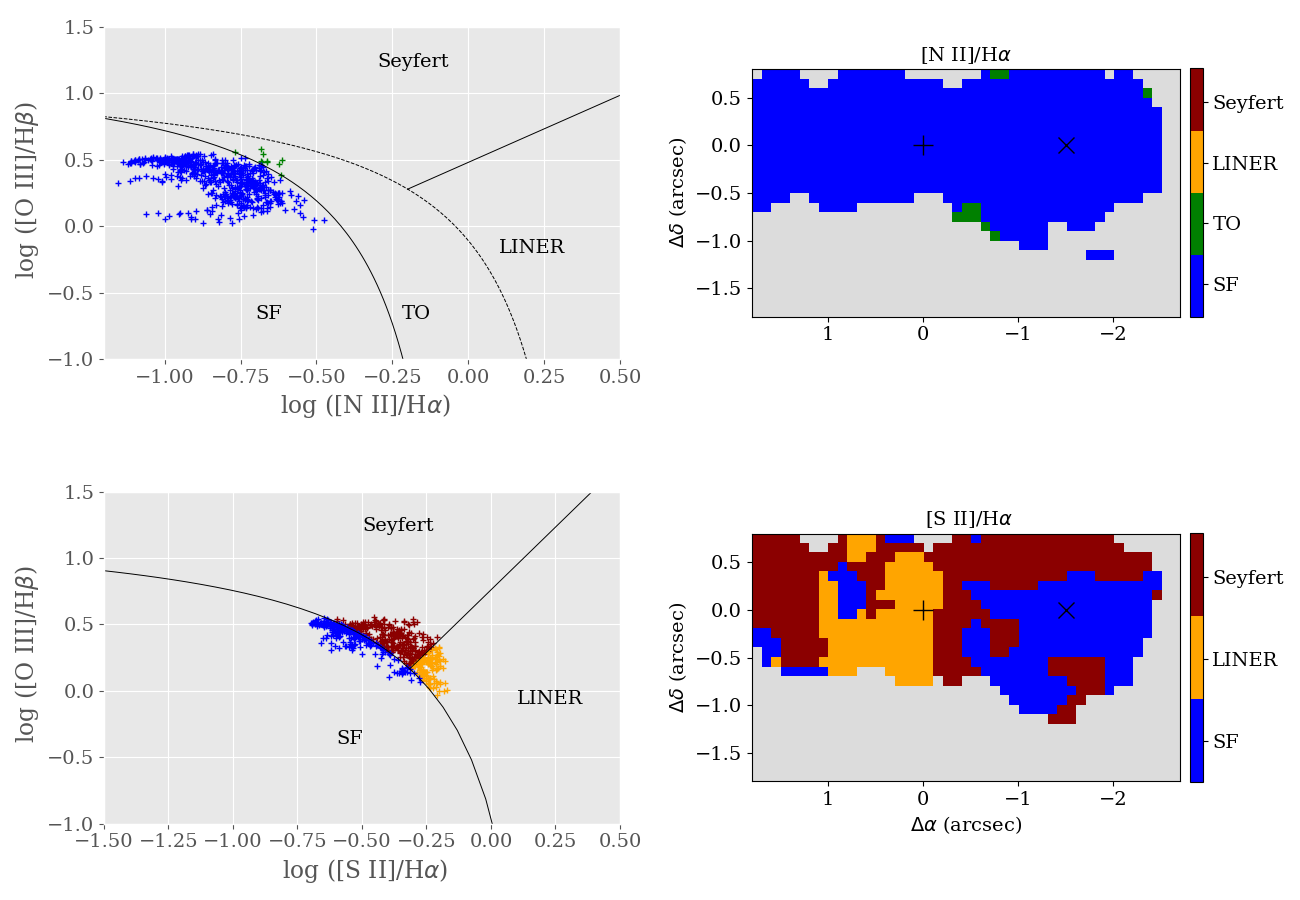}
    \caption{Optical BPT diagnostic diagram \citep{bpt_1981} for the SDSS J033549-00391 galaxy. Left upper panel: BPT -- \textbf{[N\,II]} diagram. Right upper panel: Spatially resolved BPT -- \textbf{[N\,II]} excitation map.
    Left lower panel: BPT -- \textbf{[S\,II]} diagram. Right lower panel: Spatially resolved BPT -- \textbf{[S\,II]} excitation map. }
    \label{fig:bpt_j033549}
\end{figure*}

\begin{figure*}
    \centering
    \includegraphics[scale=0.4]{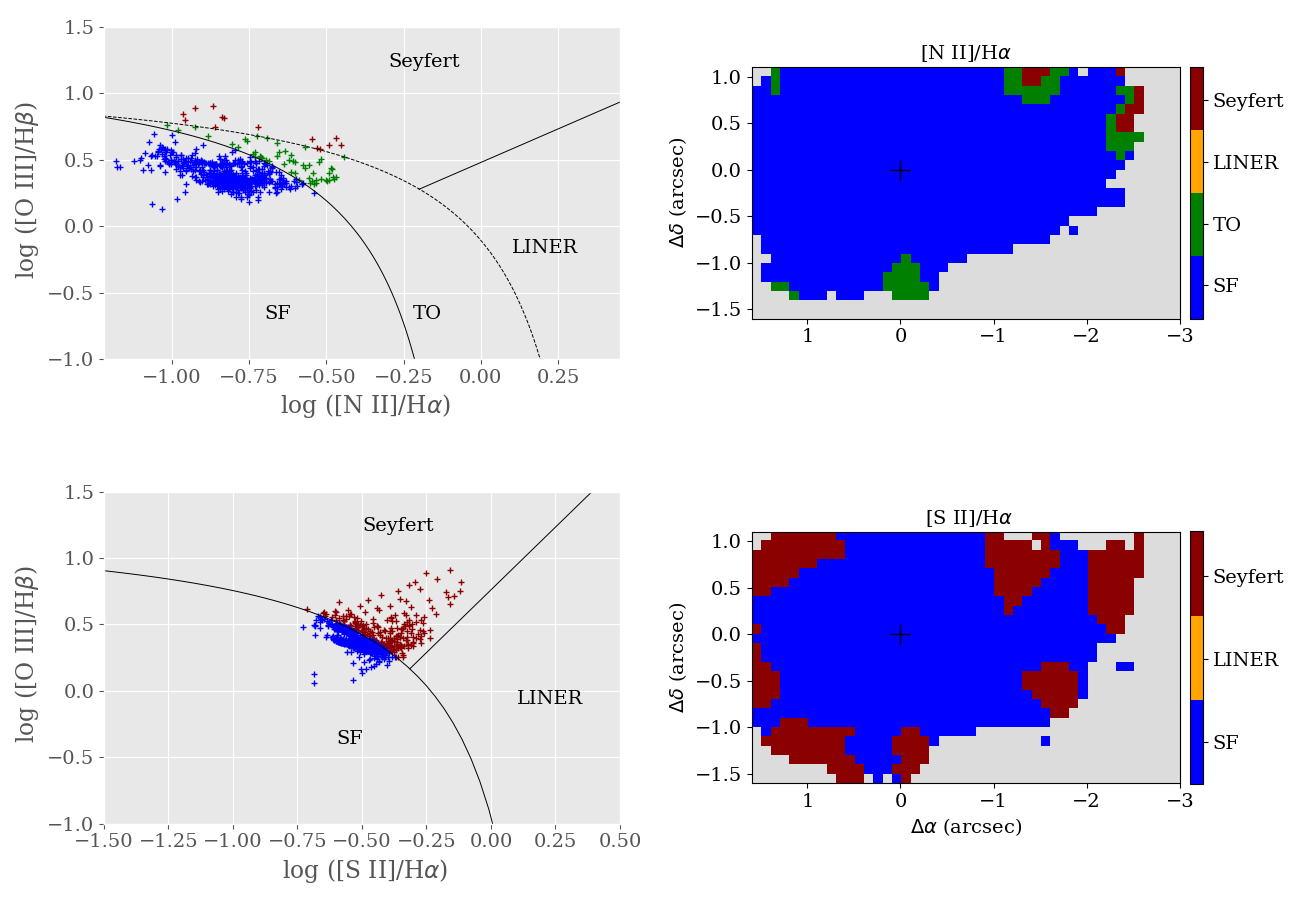}
    \caption{Same as Fig. \ref{fig:bpt_j033549} but for the SDSS J033553-003946 galaxy.}
    \label{fig:bpt_j033553}
\end{figure*}

\begin{figure*}
    \centering
    \includegraphics[scale=0.4]{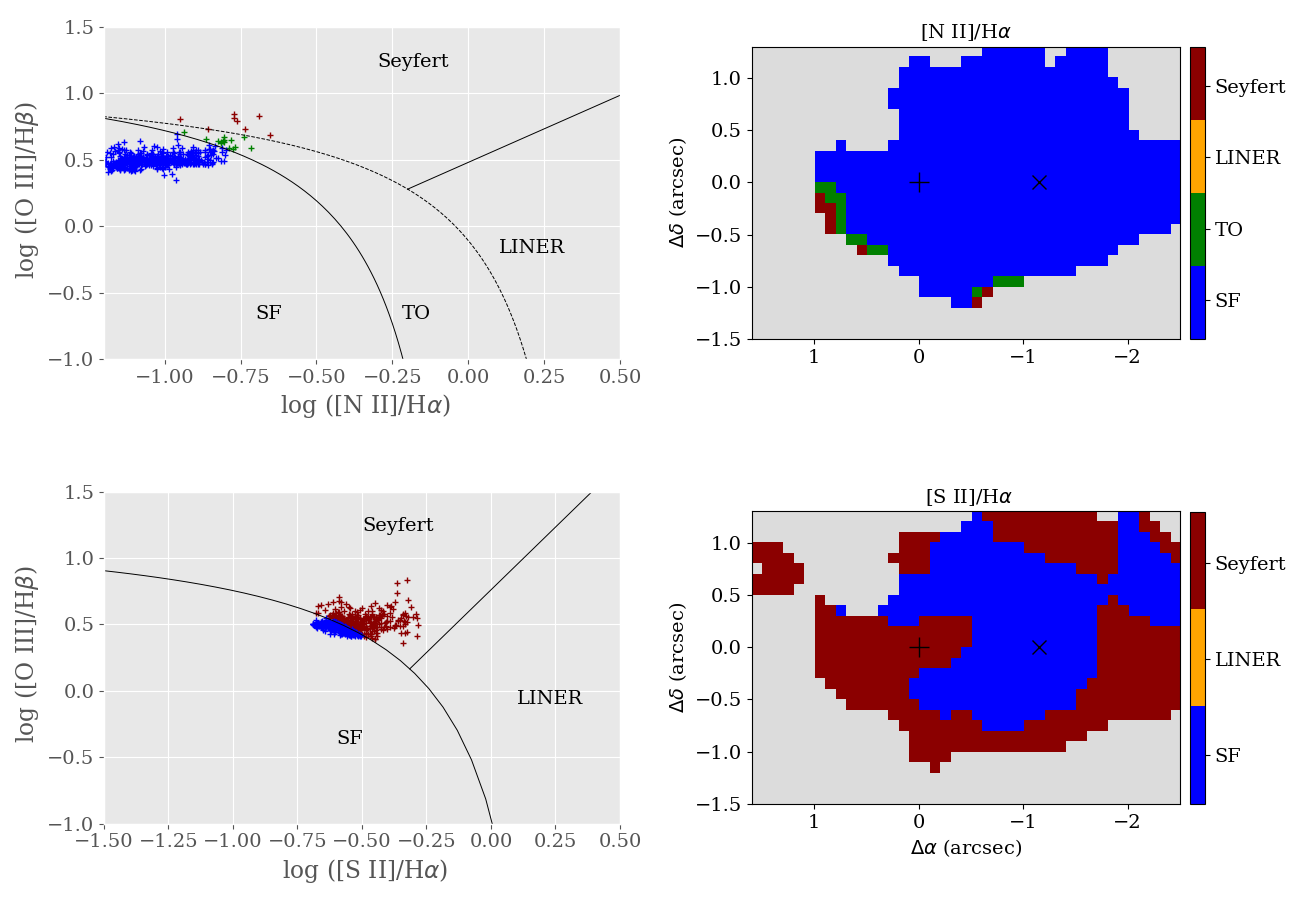}
    \caption{Same as Fig. \ref{fig:bpt_j033549} but for the SDSS J225236-00331 galaxy.}
    \label{fig:bpt_j225236}
\end{figure*}

Figures~\ref{fig:bpt_j030903} -- \ref{fig:bpt_j225236} show the BPT diagnostic diagrams and spatially resolved excitation maps for the four galaxies.  

\bsp	
\label{lastpage}
\end{document}